\documentclass[%
 reprint,
 amsmath,amssymb,
 aps,
pra,
superscriptaddress]{revtex4-2}
\usepackage{xcolor}
\usepackage{graphicx}% Include figure files
\usepackage{dcolumn}% Align table columns on decimal point
\usepackage{bm}% bold math
\begin{document}

\title{Vortex lattices in coupled one-dimensional Bose-Einstein condensates with a synthetic magnetic field}
\author{Holly A. J. Middleton-Spencer}
\affiliation{School of Physics and Astronomy, University of Birmingham, Edgbaston, Birmingham B15 2TT, United Kingdom.}
\author{Rose Davies}
\affiliation{School of Engineering \& Innovation, Aston University, 
Birmingham B4 7ET, United Kingdom.
}
\author{David G. Reid}
\affiliation{School of Physics and Astronomy, University of Birmingham, Edgbaston, Birmingham B15 2TT, United Kingdom.}
\author{Anna Dalmasso}
\affiliation{School of Physics and Astronomy, University of Birmingham, Edgbaston, Birmingham B15 2TT, United Kingdom.}
\affiliation{School of Physics and Astronomy,
The University of Nottingham, University Park, Nottingham NG7 2RD, United Kingdom.}
\author{Theodore Kazanas}
\affiliation{School of Physics and Astronomy, University of Birmingham, Edgbaston, Birmingham B15 2TT, United Kingdom.}
\affiliation{Department of Maths, University of Bath,
Claverton Down, Bath, BA2 7AY United Kingdom.}
  \author{Hannah M. Price}
 \affiliation{School of Physics and Astronomy, University of Birmingham, Edgbaston, Birmingham B15 2TT, United Kingdom.}
\date{\today}

\begin{abstract}
We investigate the mean-field ground states of coupled one-dimensional Bose-Einstein condensates subject to a synthetic magnetic field. The resulting interacting coupled-wire model has one continuous and one discrete spatial direction, providing a controlled way to interpolate between the physics of few-leg ladders and extended vortex lattices. For two wires, we study the vortex-like, biased-density, and Meissner-like states,  exploring how the finite longitudinal size of the system modifies the transitions between them. Increasing the number of wires, the ground state evolves towards an extended vortex lattice. We find numerically that periodic boundary conditions in the discrete direction can favour staggered arrays of like-signed vortices resembling an Abrikosov lattice, while open boundaries in small finite-size systems confine the vortices into rows near the centre of the synthetic direction. Our results explore how finite size effects and boundary conditions govern the emergence and spatial organisation of vortices in continuous-discrete quantum fluids, with relevance to future experiments in tunnel-coupled atomic wires or with synthetic dimensions.

\end{abstract}

\maketitle
\section{Introduction}

Quantized vortices have long been a cornerstone of cold atomic physics in two or more dimensions~\cite{abo2001observation,tsubota2002vortex,tung2006observation,cooper2008rapidly,seman2010three,williams2010observation,rosenbusch2002dynamics,aftalion2003three,kasamatsu2005three,serafini2017vortex,mccanna2021superfluid,mccanna2024curved, mccanna2024curved2, middleton2024interactions}. In a bulk gas, these vortices are typically characterized by a net phase winding around a density dip within the condensate. Under sufficiently rapid rotation or large artificial magnetic field, many such vortices organise into an extended triangular Abrikosov lattice~\cite{ tsubota2002vortex, abo2001observation,cooper2008rapidly,cooper2005vortex}. In discrete lattice systems, vortices are instead identified with circulating currents around elementary plaquettes~\cite{goldman2016topological,schafer2020tools,leonard2023realization}. The corresponding vortex arrays exhibit a rich phenomenology governed by the geometry and (magnetic-flux) commensurability of the underlying lattice~\cite{powell2011bogoliubov,reid2026phases}. 

``Coupled-wire systems" occupy the gap between a fully continuous bulk gas and a fully discrete lattice~\cite{PhysRevLett.88.036401, budich2017coupled}. In such models, particles can both move continuously along a set of parallel one-dimensional wires, and also tunnel between neighbouring wires, exploring a hybrid discrete-continuous geometry. Vortex structures in such a system are therefore also intrinsically hybrid; the density and phase of the condensate wavefunction varies continuously along the wires, but can jump discretely between the wires~\cite{chalopin2020probing}.  

Such coupled-wire constructions first came to prominence as a theoretical framework for understanding fractional quantum Hall states and other strongly-interacting topological phases~\cite{PhysRevLett.88.036401,PhysRevB.89.085101, PhysRevB.99.035130,meng2015coupled, PhysRevB.91.245139,Oreg2014Helical, Oreg2019Fractional, meng2020coupled}. However, more recently, these hybrid geometries have become experimentally relevant through several complementary routes. Firstly, coherent coupling and many-body dynamics have been studied in tunnel-coupled one-dimensional Bose gases~\cite{hofferberth2007non,chen2011many}, while arrays of closely spaced atomic wires with laser-assisted, spatially dependent tunnelling have been proposed as a way to directly implement quantum Hall coupled-wire models~\cite{budich2017coupled}. Secondly, internal or motional atomic states can be coupled together and reinterpreted as discrete sites along a synthetic dimension~\cite{celi2014synthetic, mancini2015observation,stuhl2015visualizing,ozawa2019topological,yu2025comprehensive,kanungo2022realizing,chen2024strongly, raghuram2026probing}; this leads to hybrid systems when particles can also move along extended real spatial dimensions~\cite{celi2014synthetic,price2017synthetic,salerno2019quantized,PhysRevA.111.033301}. This approach has already been employed experimentally to realize both 2D and 4D non-interacting synthetic quantum Hall coupled wire models with ultracold atoms~\cite{chalopin2020probing,li2022bec,bouhiron2024realization}. 

While the single-particle physics of coupled-wire systems is now well established~\cite{budich2017coupled, chalopin2020probing,reid2026phases, price2020synthetic, oliver2023artificial}, the interacting physics of Bose-Einstein condensates in hybrid continuous-discrete geometries is comparatively less understood~\cite{chalopin2020probing,ozawa2021artificial,ozawa2017synthetic}. In the limit of a two-wire quantum Hall system, the model becomes the same as that of a two-component spin-orbit-coupled gas, which is known to support stripe, plane-wave, and zero-momentum phases~\cite{li2012tricriticality,li2013superstripes,li2017stripe, qiu2021stripe}. As we discuss, these states are analogous, respectively, to the vortex-like, biased-density, and Meissner-like states of a fully-discrete two-leg ladder with an artificial magnetic field~\cite{Wei2014TwoLegBosons, atala2014observation, piraud2015vortex, kelecs2015mott,qiao2021quantum,uchino2015population,natu2015bosons,kolovsky2017bogoliubov}. In the opposite limit of a large hybrid quantum Hall model, a mean-field calculation has also previously shown that interactions can lead to Abrikosov-like vortex-lattice ground states ~\cite{chalopin2020probing}. However, it is not yet clear how the mean-field ground state evolves between the few- and many-wire case, and how this evolution is shaped by finite size effects and boundary conditions.

In this work, we investigate the interacting ground states of coupled one-dimensional Bose-Einstein condensates subject to a synthetic magnetic field. For two wires, we analytically and numerically recover the three phases known from spin-orbit-coupled condensates and recast these in the coupled-wire language of vortex patterns, density imbalances, and chiral currents. We then determine how transitions between these states are modified by the finite periodic length of the wires, deriving a parity-dependent quantisation of the allowed condensate momenta. By increasing the number of wires, we then explore numerically how the few-wire current patterns develops into extended vortex lattices and how these structures are reorganised by periodic or open boundary conditions in the discrete direction. Our findings establish a basis for future experiments in coupled atomic wires and synthetic dimensions, which may lead on to explorations of long-range interactions and strongly-correlated regimes.

\subsection{Outline}

We begin in Section~\ref{subsec:single_particle} by reviewing the non-interacting coupled-wire model~\cite{budich2017coupled}, before introducing its mean-field Gross-Pitaevskii description and our analytical and numerical methods in Section~\ref{subsec:meanfield}. Section~\ref{sec:2wireResults} considers the interacting two-wire system and its finite-size phase structure, while Section~\ref{sec:large} examines the emergence and boundary dependence of vortex lattices as the number of wires is increased.

\section{Mathematical Setup}\label{Sec:mathematical_setup}

As introduced above, coupled wire models originated as a novel mathematical formalism which could simply describe aspects of fractional quantum Hall physics and reproduce the hierarchy of Abelian fractional quantum Hall states \cite{PhysRevLett.88.036401}. This technique has proven a versatile approach which has since been extended to non-Abelian quantum Hall states \cite{PhysRevB.89.085101, PhysRevB.99.035130}, chiral spin liquids \cite{meng2015coupled, PhysRevB.91.245139}, fractional helical liquids \cite{Oreg2014Helical, Oreg2019Fractional} and other topological phenomena \cite{meng2020coupled}. Despite the initial mathematical motivation, this model has also recently been experimentally implemented in ultracold atomic gases to study of quantum Hall physics~\cite{chalopin2020probing,li2022bec,bouhiron2024realization}. Further theoretical proposals of experiments in cold atoms \cite{PhysRevA.111.033301, budich2017coupled} and photonics \cite{doi:10.1126/sciadv.adj0360} promise future avenues to study topological physics in such models.

\subsection{Single-Particle Coupled-Wire Model}
\label{subsec:single_particle}
\begin{figure}
    \centering
    {\includegraphics[trim={0.0cm 0.0cm 0.0cm 0.0cm},clip,width=0.9\linewidth]{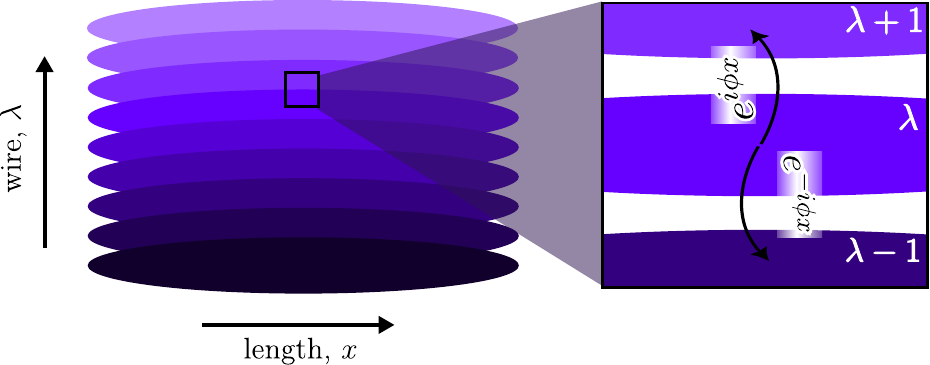}}
    \caption{Schematic of the non-interacting coupled-wire model. Particles can move freely along each wire in the (continuous) $x$ direction, and hop between wires in the (discrete) $\lambda$ direction. A synthetic magnetic flux through the $x-\lambda$ plane is encoded via the phase $\phi$, which is acquired when particles hop between neighbouring wires.}
    \label{fig:coupled_condensate}
\end{figure}

The non-interacting coupled-wire model is sketched schematically in Fig.~\ref{fig:coupled_condensate} and is described \cite{PhysRevLett.88.036401,budich2017coupled} for a particle with mass $m$, by the Hamiltonian 
\begin{equation}
      H = \int d x \sum_\lambda\left[ \psi^\dagger_{x , \lambda} \frac{\hat{p}_x^2}{2m} \psi_{x , \lambda}  - J (e^{i \phi x} {\psi}^\dagger_{x,\lambda}{\psi}_{x,\lambda+1}+ \text{h.c} ) \right]
        \label{eq:Landau}
\end{equation}
with $q=1$ and where  $\psi_{x , \lambda} $ ($\psi^\dagger_{x , \lambda} $) annihilates (creates) the particle at position ${\bf r} =(x, 
\lambda)$ with respect to the continuous and discrete directions respectively. We index the wires as $\lambda\! =\! 1,\dots, N$, where $N$ is the total number of wires. Along the discrete direction, $J$ denotes the hopping amplitude, while $\phi$ encodes a homogeneous synthetic magnetic flux through the $x\!-\!\lambda$ plane. The 1D characteristic magnetic length associated with this flux is then given by $l_B = 2 \pi /\phi = 1/B$, which sets a natural length-scale of interest along the continuous direction. Note that throughout we have set the lattice spacing along the discrete direction equal to one.
The single-particle physics of Eq.~\ref{eq:Landau} has previously been studied both experimentally~\cite{chalopin2020probing} and theoretically~\cite{budich2017coupled}.

If periodic boundary conditions are applied along the discrete direction then the corresponding motion can be expressed in terms of plane-wave states with $N$ different quasi-momenta $k_\lambda \!\in\! [ -\pi, \pi)$. For a given $k_\lambda$, the first-quantized Hamiltonian along $x$ is then given by~\cite{budich2017coupled}
\begin{equation}
h_{x, k_\lambda} = -\frac{\hbar^2}{2m} \frac{\partial^2}{\partial x^2} - 2 J \cos (\phi x + k_\lambda), \label{eq:h1d} 
\end{equation}
which corresponds to a 1D Hamiltonian describing a particle moving in a oscillatory potential with a period given by the magnetic length $\ell_B$. This can be solved using standard techniques by applying periodic boundary conditions along the continuous $x$ direction and introducing the quasi-momenta $k_x \in [ -\phi/2, \phi/2)$ to find the energy band-structure~\cite{budich2017coupled}. Importantly, the resulting energy bands are independent of $k_\lambda,$ which only shifts the phase of the potential in Eq.~\ref{eq:h1d}, and therefore there is an $N$-fold degeneracy at every $k_x$ value. This degeneracy is guaranteed by the periodic boundary conditions in the discrete direction. These energy bands can also each be associated with a topological Chern number, which is equal to one for low-lying bands, irrespective of the parameters - so long as the band gaps remain open~\cite{budich2017coupled}. 

\begin{figure}
    \centering
    {\includegraphics[trim={0.0cm 0.0cm 0.0cm 0.0cm},clip,width=0.9\linewidth]{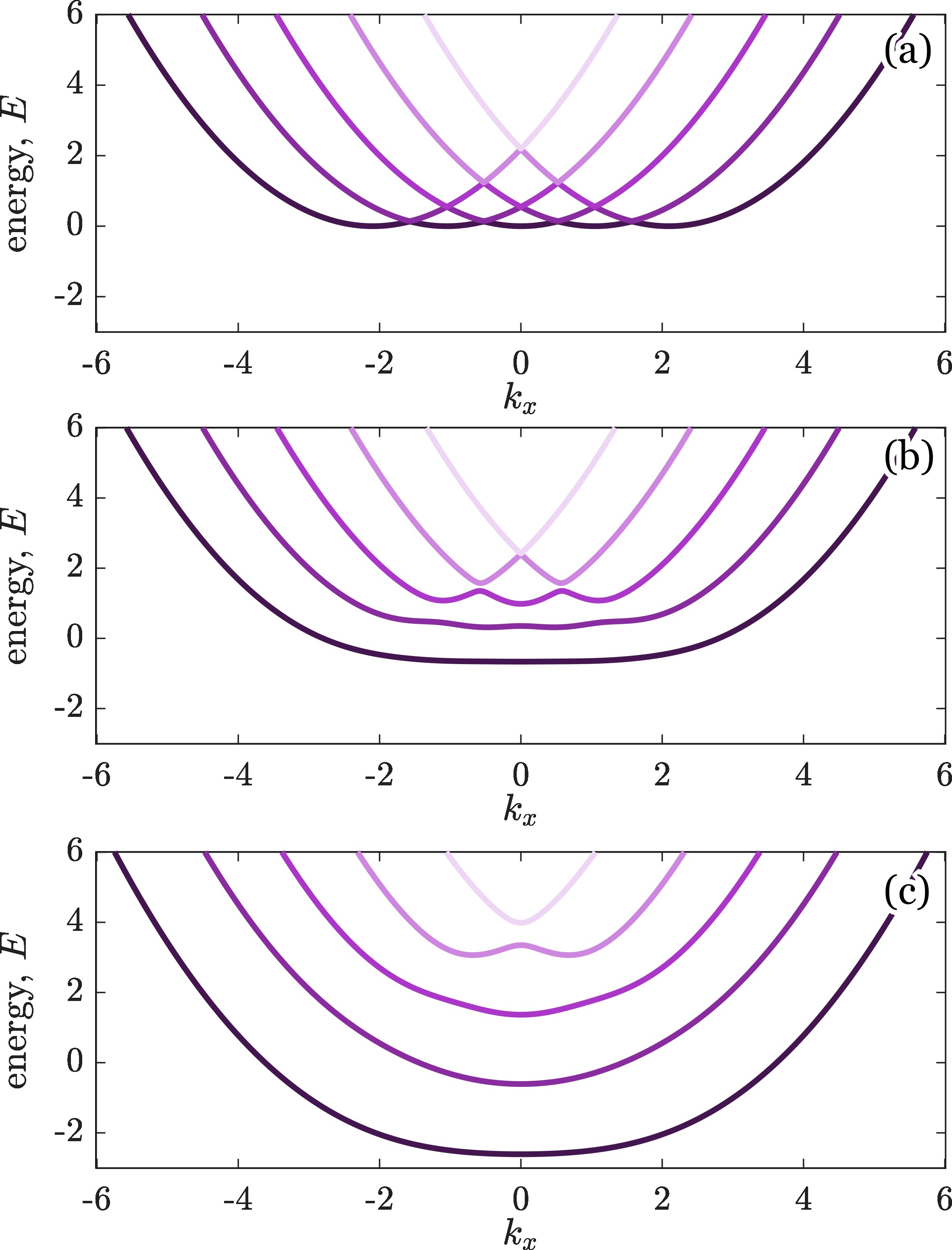}}
    \caption{The band structure for a non-interacting coupled wire system with $N=5$ wires and open boundary conditions in the discrete direction and periodic boundary conditions in the continuous direction with $\phi=2\pi/6$, and (a) $J = 0.01$, (b) $J = 0.6$ and (c) $J=1.8$. Here $\hbar=m=1$.}
    \label{fig:band_structure}
\end{figure}

Another instructive way to solve the single-particle coupled-wire model is to impose open boundary conditions along the discrete direction, while maintaining periodic boundary conditions in the continuous direction. In this case, it is natural to apply a gauge transformation to Eq.~\ref{eq:Landau} in order to write
\begin{align}\label{eq:OBCHam}
     H = \int d x \sum_\lambda &\biggl[ \tilde{\psi}^\dagger_{x , \lambda} \frac{\hbar^2}{2m}\big(-i\partial_x + \phi_\lambda\big)^2 \tilde{\psi}_{x , \lambda}  \\
     & - J({\tilde{\psi}}^\dagger_{x,\lambda}{\tilde{\psi}}_{x,\lambda+1}+ \text{h.c} )\biggr] \,, \nonumber
 \end{align}
where $\phi_\lambda = 
\frac{(N-1)\phi}{2} - \phi(\lambda-1)$, which can straightforwardly be diagonalised. The resulting spectrum is shown for an exemplar non-interacting five wire system in Fig.~\ref{fig:band_structure} for $\phi=2\pi/6$, and (a) $J = 0.01$, (b) $J = 0.6$ and (c) $J=1.8$. As can be seen, for small values of  $J/\phi^2$ [c.f. panel~(a)], the lowest energy band still consists of $N$ distinct minima in the thermodynamic limit. As $J/\phi^2$ is increased [c.f. panel (c)], these minima merge into a single global minimum. The analytical solution for the two-wire case will be discussed in Section~\ref{sub:two-wire-open}.

% For open boundary conditions, it is convenient for later analytics to express the non-interacting Hamiltonian with the rotated field operators, $\tilde{\psi}_{x,\lambda} = \exp(i \phi(\lambda-1) x )\psi_{x,\lambda}$, in the Fourier transformed representation 

% \begin{align}\label{eq:OBCHam}
%     H = \int d x \sum_\lambda &\left[ \tilde{\psi}^\dagger_{x , \lambda} \frac{\hbar^2}{2m}(\hat{p}_x - \phi (\lambda-1))^2 \tilde{\psi}_{x , \lambda}  \right. \\
%     &\left. - \hbar J({\tilde{\psi}}^\dagger_{x,\lambda}{\tilde{\psi}}_{x,\lambda+1}+ \text{h.c} )\right] \,, \nonumber
% \end{align}
% as this form allows the spectrum and eigenstates to be easily found through diagonalising. The resulting spectrum consists of $N$ bands but without the guarantee of degeneracy at each value of $k_x$. For small values of  $J/\phi^2$, the lowest energy band consists of $N$ minima [c.f. Fig.~\ref{fig:open-2-wire-Var} (c) \textit{red line} for the two wire case]. Notably, these minima eventually merge into one global minimum as $J/\phi^2$ is increased (c.f. Fig.~\ref{fig:open-2-wire-Var} d) \textit{cyan line} for two wire case). The analytical solution for the two-wire case will be discussed in Section~\ref{sub:two-wire-open}.

\subsection{Mean-Field Bosonic Coupled-Wire Model}\label{subsec:meanfield}

By including mean-field bosonic contact interactions, one can derive a set of coupled mean-field Gross-Pitaevskii equations from Eq.~(\ref{eq:Landau}) as
\begin{equation}
    i\frac{\partial\psi_\lambda}{\partial t} = -\frac{1}{2}\frac{\partial^2 \psi_\lambda}{\partial x^2} - J[e^{i\phi x}\psi_{\lambda+1}+e^{-i\phi x}\psi_{\lambda-1}]+|\psi_\lambda|^2\psi_\lambda,
    \label{eq:gpe_1}
\end{equation}
where $\psi_\lambda (x)$ is the mean-field condensate wave-function associated with the wire indexed by $\lambda$. Note that Eq.~\ref{eq:gpe_1} has been rescaled such that $x\rightarrow x\xi$, where $\xi$ is the healing length of the system, $\xi=\hbar/\sqrt{m\mu}$, $J\rightarrow J'\mu$, and $t \rightarrow t'\hbar/\mu$, where $m$ is the mass of the atom, and $\mu$ is the chemical potential of the system which is set to be $g \rho_{\lambda=1}$, with $\rho_{\lambda=1}$ being the peak density of the first wire. %This scaling is typical for 2D vortices.

We solve Eq.~\ref{eq:gpe_1} using two complementary approaches. Firstly, following the well-established tradition of introducing a variational ansatz in interacting quantum systems \cite{PhysRevLett.50.1395, doi:10.1126/science.235.4793.1196, Clark_2006, PhysRevB.101.045102}, we consider
\begin{equation}
\Psi (x, \lambda) = \sum_j a_j \varphi_j (x, \lambda)\,, \label{eq:ansatz}
\end{equation}
where each $a_j$ is a variational complex coefficient, obeying the normalisation condition, $\sum_j \vert a_j\vert^2=1$, and where $\varphi_j (x, \lambda)$ denotes the single-particle wave-function associated with the $j$-th energy minimum of the lowest band, as described further below; note that this ansatz is only valid in the weakly interacting regime.

In the case of periodic boundary conditions in both directions (c.f. Eq.~\ref{eq:h1d}), the bands for each value of $k_x$ are $N$-fold degenerate, as previously mentioned. At the same time, for a given $k_\lambda$, there is only one minimum in the lowest energy band, which occurs at $k_x \!=\! 0$. Therefore in Eq.~\ref{eq:ansatz}, there are $N$ possible $\varphi_j (x, \lambda)$ states to be included in the ansatz, such that $j=1,...,N$. As each of these states already minimizes the single-particle kinetic energy, all that remains is to minimize the interaction energy $E_{\rm{int}}=  g\int_0^L dx|\Psi|^4$ with respect to the set of $a_j$ (subject to normalization) to approximate the ground-state wavefunction. We perform this minimisation analytically in the two-wire case and numerically for more wires by using an in-built minimiser following the quasi-Newton method of Broyden, Fletcher, Goldfarb, and Shanno (BFGS) \cite{10.1093/comjnl/13.3.317} with the initial values of $a_j$ determined from a pseudo-random uniform distribution. Note that as the number of complex coefficients scales with the number of wires, this procedure becomes more challenging as $N$ increases. 

If instead we consider open boundary conditions along the discrete direction, there is no longer a guaranteed degeneracy in the lowest energy state as a function of $k_x$ as can be seen in Fig.~\ref{fig:coupled_condensate}. To reflect this, we make the number of single-particle wavefunctions to be included in the variational ansatz parameter dependent. For parameters which result in only one minima the minimisation procedure is trivial. These solutions are associated with non-vortex states \cite{Wei2014TwoLegBosons, li2012tricriticality}, which will be discussed later. Note that a final subtlety in this analysis stems from the interplay of the length of the wires, $L$, and the 1D magnetic length, $2\pi/\phi$. These scales must be commensurate to guarantee that the hopping term in Eq.~\ref{eq:Landau} is periodic along the wire, as discussed further below. Overall the condensate solution in each wire must have a well defined phase - in superconductors this requirement leads to the well known quantisation of flux \cite{byers1961theoretical}. This restriction can result in the condensate solution not corresponding to the `true' minima of the spectrum depending on the chosen wire length $L$. This will be explored analytically in Section \ref{sub:two-wire-open}.

The second numerical approach that we will use is the standard technique of imaginary time evolution to find the ground state of the mean-field coupled wire model. In this approach, we solve Eq.~\ref{eq:gpe_1} using a fourth-order Runge-Kutta method in imaginary time $t\rightarrow-it$, monitoring the density at every time-step. We evolve the system until the variation in total density between time-steps is given by $\Delta n\sim\mathcal{O}(10^{-14})$. At each step, the peak density of the wavefunction is normalised to one. 

These two separate numerical methods complement each other well. In particular, the variational technique has two immediate advantages when compared to imaginary time evolution; firstly, the resulting ground-states are related to a few modes from the underlying single-particle band structure, providing a clean physical intuition. Secondly, due to the reduced parameter-space the variational method is comparatively computationally inexpensive. However, the variational method is generally less accurate because it makes the assumption that only a few modes contribute, leading to the method breaking down at a smaller $g$ than where the imaginary time evolution method fails. 

The ground states found can be defined and categorised by a few methods. Firstly, one can simply look at the density, $\rho_\lambda=|\psi_\lambda|^2$ and the phase $\theta_\lambda$ of the condensate $\psi_\lambda= \sqrt{\rho_\lambda}e^{i\theta_{\lambda}}$. In a typical continuous-space mean-field condensate, a vortex consists of a density dip, with an associated $2\pi$ phase winding around this point, and can therefore be identified by inspection from plots of these two quantities. Conversely, in lattice setups, vortices are typically quantified by looking at currents between individual lattice sites. To this end, we define the intra-wire current (along the continuous direction) as
\begin{equation}
    {j}_{\lambda}^{\parallel}(x) = -\frac{i}{2}\bigg[\psi^{*}_\lambda\frac{\partial\psi_\lambda}{\partial x}-\frac{\partial\psi_\lambda^*}{\partial x}\psi_\lambda\bigg] =  \rho_\lambda \frac{\partial\theta_\lambda}{\partial x}\,.
    \label{eq:along_current_j}
\end{equation}
We can also define the current associated with the hopping between each wire at a given location, $x$,
\begin{equation}
    {j}^{\perp}_{\lambda,\lambda+1}(x) = -iJ[e^{i\phi x}\psi_\lambda^*\psi_{\lambda+1}-e^{-i\phi  x}\psi_{\lambda}\psi^*_{\lambda+1}]
    \label{eq:inter_current_j}.
\end{equation}
Note that the above two equations are the appropriate expressions for the gauge choice in Eq.~\ref{eq:Landau}, which is used throughout for the imaginary time method. If we instead use the gauge choice of Eq.~\ref{eq:OBCHam}, then these should be modified to be,
\begin{align}
    \tilde{j}_{\lambda}^{\parallel}(x) &= \tilde{\rho}_\lambda \left(\frac{\partial\tilde{\theta}_\lambda}{\partial x} + \phi_\lambda \right) \,,\\
    \tilde{j}^{\perp}_{\lambda,\lambda+1}(x) &= -iJ[\psi_\lambda^*\psi_{\lambda+1}-\psi_{\lambda}\psi^*_{\lambda+1}]\, ,
\end{align}
where now we have used that $\tilde{\psi}_{\lambda}= \sqrt{\tilde{\rho}_{\lambda}}e^{i\tilde{\theta}_{\lambda}}$. 

\section{Interacting States in Two coupled wires}\label{sec:2wireResults}

In this section, we focus on the case of two coupled wires. This provides the simplest geometry in which to analyse the interacting states of the coupled-wire model. We first consider periodic boundary conditions along the discrete direction, which give a minimal closed geometry. We then turn to open boundary conditions along the discrete direction, where the model supports vortex-like, biased-density, and Meissner-like phases, similar to in the two-leg fully-discrete Hofstadter ladder~\cite{atala2014observation, piraud2015vortex, kelecs2015mott,qiao2021quantum,uchino2015population,natu2015bosons,kolovsky2017bogoliubov,reid2026phases}. In this limit, the coupled-wire model is equivalent to a variant of the previously-studied spin-orbit-coupled two-component Bose gas, where these phases are known as the stripe, plane-wave and zero-momentum phases, respectively~\cite{li2012tricriticality}. We connect these two pictures and highlight how these phases can be characterized in terms of vortex patterns, density imbalances and chiral currents. We also study finite-size effects, showing how the requirement of periodicity along the wire imposes a holonomy constraint on the allowed condensate momenta, leading to parity-dependent finite-size phase sequences.

\begin{figure}
    \centering
    \includegraphics[trim={0.0cm 0.0cm 0.0cm 0.0cm},clip,width=0.9\linewidth]{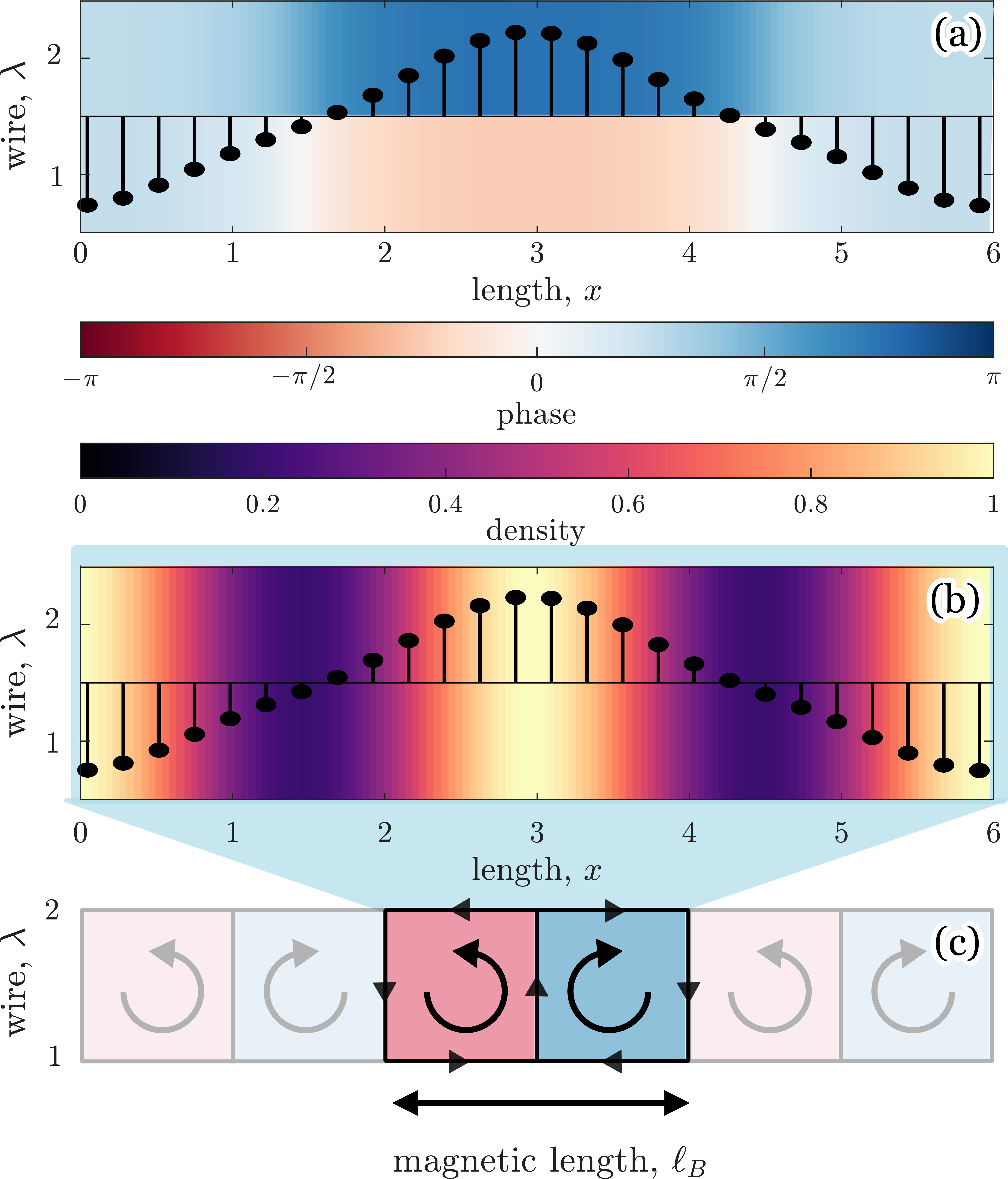}
    \caption{(a) The phase and (b) density of the interacting ground state of two coupled wires with periodic boundaries conditions in both directions as obtained from solving Eq. \ref{eq:gpe_1} via imaginary time evolution for $\phi=2\pi/6$, $\ell_B\!=\!6$, $L=l_B$ and $J\!=1\!$. The inter-wire current ${j}_{1,2}^\perp$ (Eq. \ref{eq:inter_current_j}) is superimposed as black dots on each figure (with an arbitrary scale) to better see the correspondence between density, phase and inter-wire current. (c) Schematic of the ground state as a vortex-antivortex pair per magnetic length, with arrows and shading indicating the direction of current flow.}
    \label{fig:alpha_1-6_N2}
\end{figure}

\subsection{Periodic discrete boundaries}
\label{sub:two-wire-per}

We begin by analysing the interacting ground state of two coupled wires with periodic boundary conditions along both the continuous and discrete directions. Note that when imposing periodic boundary conditions along the continuous direction, we will always assume that the length of the wires, $L$, is an integer multiple of the magnetic length, i.e. $L=z\ell_B$, $z\in \mathbb{N}$, to ensure the hopping phase-factor, $e^{i\phi x}$  in Eq.~\ref{eq:gpe_1}, is single-valued.

For two wires, imposing periodic boundary conditions in the discrete direction should be understood as taking a two-site ring which includes two distinct links between the same two wires. These two links have complex-conjugated position-dependent hopping amplitudes, which combine to produce an effective total hopping $2J \cos(\phi x)$ between the two sites. The single-particle
problem can then be diagonalised by introducing the bonding and antibonding
fields in the usual way for a two-site model as
\begin{equation}
\psi_\pm(x)=\frac{1}{\sqrt{2}}\left[\psi_1(x)\pm\psi_2(x)\right].
\end{equation}
In these variables, the non-interacting Schr\"{o}dinger equation reduces to a pair of
1D equations,
\begin{equation}
\left[
-\frac{1}{2}\frac{d^2}{dx^2}
\mp 2J\cos(\phi x)
\right]\psi_\pm(x)
=
\epsilon \psi_\pm(x),
\end{equation}
with energy $\epsilon$,  where the sign depends on the bonding or antibonding sector. 
Thus a single particle experiences an effective periodic potential controlled
by the inter-wire hopping, with period $\ell_B=2\pi/\phi$. The
single-particle eigenstates are Mathieu functions, and the density
profile inherits this magnetic-length periodicity.

The effective-potential picture also gives a simple interpretation of
the density modulation. Near a minimum of the lower effective potential,
\begin{equation}
-2J\cos(\phi x)
\simeq
-2J+J\phi^2(x-x_0)^2,
\end{equation}
so the local dynamics are approximately those of a harmonic oscillator
with frequency
\begin{equation}
\omega=\phi\sqrt{2J}.
\end{equation}
The characteristic width of the density peak therefore scales as
$\sigma\sim(2J\phi^2)^{-1/4}$, explaining why the density modulation
becomes more pronounced as $J$ is increased.

These single-particle features carry over to the weakly-interacting
regime; this is shown in Figure \ref{fig:alpha_1-6_N2} for $J=1$, and $\phi=2\pi/6$ for a system of length $L=\ell_B$, as determined via imaginary time evolution. (The same qualitative state is obtained with the variational method.) As expected, the phase [panel (a)] and density [panel (b)] are both periodic with the magnetic length, with two density dips occurring on each wire concurrently with a sharp increase or decrease in the phase. The direction of these phase changes are opposite for the different wires meaning that the intra-wire current (Eq.~\ref{eq:along_current_j}) is opposed between the two wires. We also superimpose the inter-wire current $j^\perp_{\lambda,\lambda+1}$ (Eq. \ref{eq:inter_current_j}) as black dots (with an arbitrary scale) on both panels (a) and (b). From this we can see that the inter-wire current varies approximately sinusoidally over the magnetic length, with maxima occurring midway between the density dips in regions of constant phase. 

The longitudinal and inter-wire currents therefore form two regions of opposite circulation within each magnetic length. We interpret these as like a vortex--antivortex pair located between the two wires, as sketched schematically in Fig.~\ref{fig:alpha_1-6_N2}(c). Since the wire direction is
continuous and the transverse direction contains only two sites, these
are not vortices in the usual lattice sense where the current winds around a discrete plaquette, but
rather an emergent coupled-wire analogue of such a vortex--antivortex current
pattern.
Similar results are found for all values of $J$ and $\phi$ studied, with the depth of the density dips increasing as $J$ increases, consistent with the effective-potential estimate above.
From the variational method, these vortex states can be understood as arising from the equal superposition of the two degenerate minima in the single-particle periodic band-structure. 

\subsection{Open discrete boundaries}

If we instead impose open boundary conditions along the discrete direction, our model is closely related to the widely-studied two-leg bosonic Hofstadter ladder~\cite{atala2014observation,reid2026phases}, in which both directions are discrete. %Similarly to that system, we are able to identify distinct phases in different parameter regimes. 
As a function of the flux and imbalance of the inter-leg and intra-leg hopping amplitudes, the two-leg Hofstadter model has a well-known phase diagram, with three distinct types of phases. These are the ``vortex'' phase, where the current forms closed plaquettes with corresponding density oscillations, as well as two phases that have uniform densities along the legs. These uniform phases are usually referred to as the  ``Meissner" (or ``saturated-chiral-current") phase, and the biased-ladder phase. In both the latter, the two legs carry oppositely-directed longitudinal currents, with the density being, respectively, equal or imbalanced between the two legs~\cite{orignac2001meissner, Wei2014TwoLegBosons}. 

This two-wire coupled-wire model is equivalent to a specific type of two-component spin-orbit-coupled Bose-Einstein condensate. In this language, the vortex-like, biased-density and Meissner-like states correspond to the stripe, plane-wave and zero-momentum phases, respectively~\cite{li2012tricriticality,li2013superstripes}. Similar spin-orbit BECs have been realised experimentally, with studies probing their phase diagram and directly observing stripe order~\cite{li2013superstripes,ji2014experimental,li2017stripe}. Unlike these experiments, our interacting model corresponds to having equal intra-component interactions and vanishing inter-component interactions in the spin-orbit coupled gas. The stripe states in this case have previously been studied and interpreted as arrays of Josephson vortices, with nonlinear stationary solutions for their density, phase, and current profiles being obtained~\cite{qiu2021stripe}. Related connections between fully discrete bosonic flux ladders and two-component spinor gases have also been discussed previously~\cite{citro2020spectral,cabedo2020effective}. Here, we recover the two-wire interacting ground states via a simple band-minimum variational ansatz and characterise these states in the coupled-wire language of longitudinal, inter-wire, and chiral currents. We also verify our results with imaginary-time evolution of the full Gross-Pitaevski equation. This provides the starting point for our analysis of finite-length momentum quantisation and of the evolution towards many-wire vortex structures.

\label{sub:two-wire-open}
\subsubsection{Variational method}
\label{sec:variational_2_wire}

\begin{figure*}
    \centering
    {\includegraphics[trim={0.0cm 0.0cm 0.0cm 0.0cm},clip,width=0.95\linewidth]{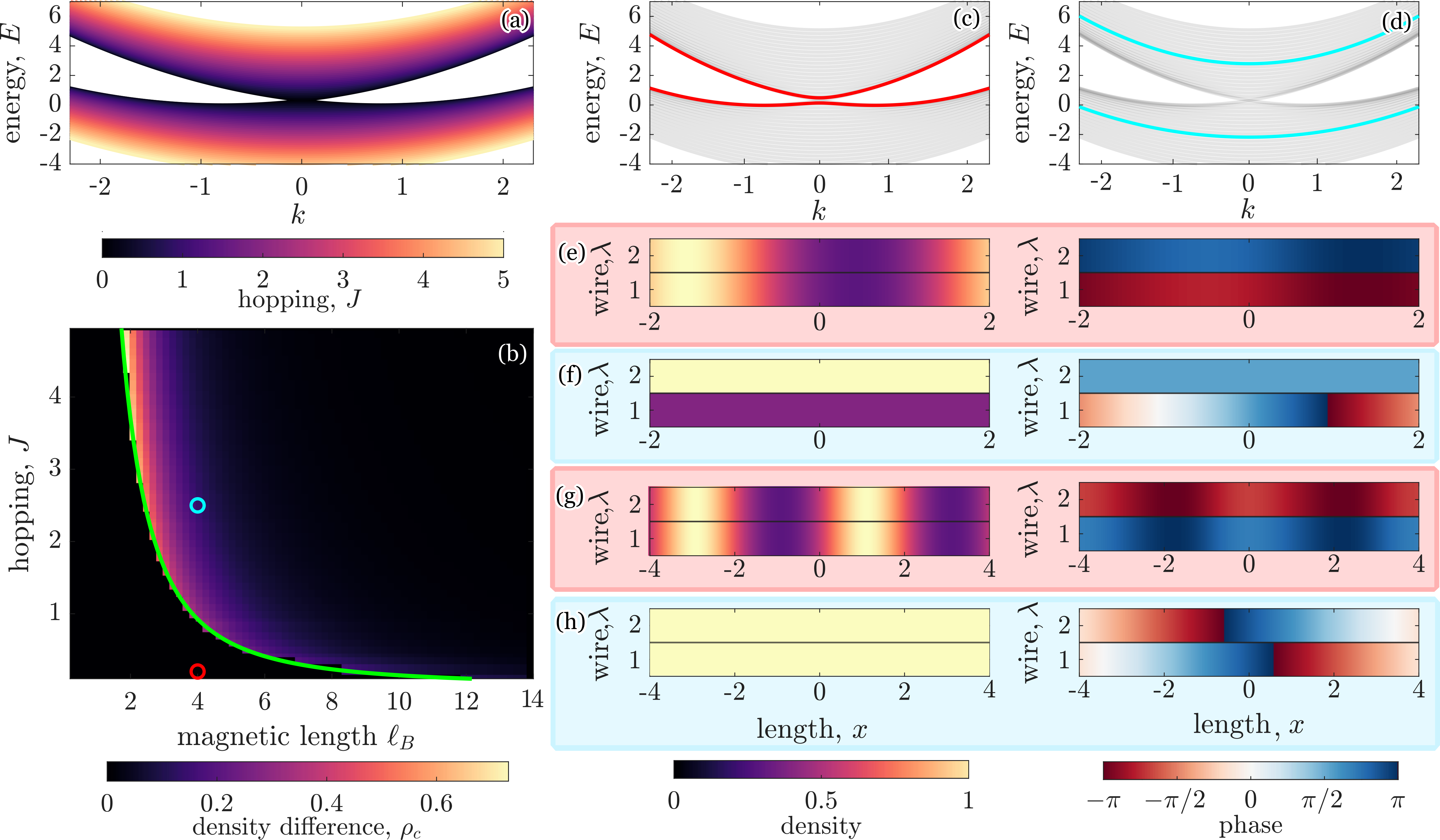}}
    \caption{Interacting ground-states of two coupled wires with open boundary conditions along the discrete direction as obtained numerically from the variational method. (a) The non-interacting band structure [Eq. \ref{eq:spectrum2wireOBE}] for hopping values ranging from $J=0.0$ to $J=5.0$ with $\phi=\pi/2$ (i.e. $\ell_B=2 \pi/\phi=4)$.
      (c) \& (d) The non-interacting energy bands  ({\it grey lines}), replotted from panel (a), highlighting (c) $J=0.2$ ({\it red line}) and (d) $J=2.5$ ({\it blue line}). For the highlighted band in (c), the lowest energy band has two minima, while in (d) the band has a single minimum at $k_x=0$.     
    (b) The numerically-calculated spatially-averaged density difference, $|\rho_c|\equiv| \bar\rho_1-\bar\rho_2|$,  plotted as a function of the magnetic length ($\ell_B=2\pi/\phi$) and the inter-wire hopping $J$ and a wire length of $L=z l_B=l_B$, with $z=1$. As $z$ is odd, finite-size effects mean that the $k_x=0$ minimum is not accessible and so the system can only be in either the vortex phase or the biased-density phase. The analytical prediction for the transition between these two phases (Eq. \ref{eq:vortex-biased-transition}) is indicated by the green line. The red and blue dots mark the same parameters as used in panels  (c) \& (d) respectively. (e) \& (f) The corresponding  density ({\it left}) and phase ({\it right}) profiles for these same parameters, showing that as expected, for (e) [resp. (f)] the system is in the vortex [resp. biased-density] phase, as  characterized by a modulated [resp. uniform] density along the wires, and vanishing [resp. non-zero] density imbalance between the wires.
    %In contrast, (d) the band structure for the density ladder case has a single minima. 
    (g) \& (h) The density (\textit{left}) and  phase (\textit{right}) profile of the ground state again for the same parameters as in (c) \& (d) respectively, except now with $L=2 \ell_B$. As can be seen, (g) is qualitatively very similar to (e), as the system is again in the vortex phase. However, (h) is now distinct from (f), as finite-size effects do not preclude the system from reaching the Meissner-like phase, corresponding to a condensate at $k_x=0$ with zero density imbalance. 
   }
    \label{fig:open-2-wire-Var}
\end{figure*}

\paragraph{Thermodynamic limit.---}
In the non-interacting case, the spectrum and associated wavefunction of eigenstates in the lower band can be directly obtained from solving Eq.~\ref{eq:OBCHam} with $N=2$. Following Ref \cite{budich2017coupled} and their choice of gauge, the spectrum and wavefunction are given in our rescaled variables as, 
\begin{align}
    \varepsilon(k) &= \frac{k^2}{2} + \frac{\phi^2}{2}\Bigg(\frac{1}{4} - \sqrt{\frac{k^2}{\phi^2} + \tilde{J}^2} \ \Bigg)\,,\label{eq:spectrum2wireOBE} \\
    \psi_k(x) &=  \frac{1}{(f_k+f_k^{-1})^{1/2}}\begin{pmatrix}
        f_k^{-1/2}e^{-i\phi x/2} \\ f_k^{1/2}e^{i\phi x/2}
    \end{pmatrix}e^{ikx}\,,\label{eq:BudichWavefunc}
\end{align}
where $k\equiv k_x$, $\tilde{J} \equiv 2J/\phi^2$ and $f_k\equiv \sqrt{1+(k/\tilde{J}\phi)^2} - k/\tilde{J}\phi$ with $f^{-1}_k = f_{-k}$. The components of the vector describe the wavefunction on each wire. Note that the spectrum is an even function of $k$ and therefore there is always an energy degeneracy between states $\psi_{\pm k}$. Generically the spectrum has two minima at $k_\pm\equiv\pm \phi/2 \sqrt{1-4\tilde{J}^2}$ for $\tilde{J}<1/2$ and a single minima at $k=0$ for $\tilde{J}\ge 1/2$. The low-energy part of the spectrum is plotted in Fig.~\ref{fig:open-2-wire-Var}(a) for $\phi=\pi/2$ and a range of hopping amplitudes from $J=0$ to $J=5$. Examples of the spectrum with double and single minima are highlighted by the red and blue lines in panels (c) for $J=0.2$, and (d) for $J=2.5$, respectively. 
For $\tilde J<1/2$, the variational ansatz in the thermodynamic limit can be
constructed from the two degenerate minima as
\begin{equation}
\psi(x)
=
a_1\psi_{k_+}(x)
+
a_2\psi_{k_-}(x),
\qquad
|a_1|^2+|a_2|^2=1 .
\end{equation}
As these states have the same single-particle energy, the variational method minimizes the interaction energy with respect to the relative occupation of the minima, through $a_1$ and $a_2$. Using
Eq.~\ref{eq:BudichWavefunc}, the interaction energy within this
two-mode subspace can be written as
\begin{equation}
\frac{E_{\rm int}}{gL}
=
\frac{f_k^2+f_k^{-2}}
{(f_k+f_k^{-1})^2}
\left(
2|a_1|^4-2|a_1|^2+1
\right)
+
\frac{8|a_1|^2(1-|a_1|^2)}
{(f_k+f_k^{-1})^2},
\label{eq:EintThermo}
\end{equation}
This interaction energy is independent of the relative phase between $a_1$ and $a_2$, which simply translates any density modulations along the wire. The polynomial is minimised
by either an equal occupation of the two momenta,
$|a_1|^2=|a_2|^2=1/2$, or by occupation of only one of the two momenta,
$|a_1|^2=0$ or $1$. The boundary between these two possibilities occurs
when
\begin{equation}
\tilde J_c
=
\sqrt{2}\frac{|k|}{\phi}=
\frac{1}{\sqrt{2}}
\sqrt{1-4\tilde J_c^2}, \quad \rightarrow \quad \tilde J_c=\frac{1}{\sqrt{6}}. 
\end{equation}
In the thermodynamic limit, the equal-superposition state is therefore
favoured for $\tilde J<1/\sqrt{6}$, corresponding to a vortex-like
phase, while the single-minimum state is favoured for
$1/\sqrt{6}<\tilde J<1/2$, corresponding to a biased-density phase. 
Finally, for $\tilde J\geq 1/2$, the lower band only has a single minimum at
$k=0$, giving the density-balanced Meissner-like phase. Note that the distinction between the vortex-like phase and biased-density phase is due to interactions, as these states are degenerate by construction for a single particle. This interaction-driven transition therefore
disappears in the non-interacting limit, and only the change from a general superposition state to the Meissner-phase at $\tilde J=1/2$ remains~\cite{budich2017coupled}. The above results are only valid for weak interactions, as the transitions will generally be shifted when interactions modify the optimal condensate momentum \cite{li2012tricriticality}. For example, in the equivalent homogeneous spin-orbit-coupled condensate, sufficiently strong spin-dependent interactions (corresponding to inter-wire interactions which were are not included in Eq.~\ref{eq:Landau}) can even eliminate the intermediate plane-wave phase, producing a vortex-like to Meissner-like state transition~\cite{li2012tricriticality}. 

Let us now describe the distinguishing properties of these three states.
We begin with the vortex-like phase; when both minima are equally
occupied, similar to in the Harper-Hofstadter ladder~\cite{atala2014observation,orignac2001meissner,Wei2014TwoLegBosons,budich2017coupled}. This state is also known as the stripe phase in a spin-orbit-coupled condensate~\cite{li2012tricriticality}, and its alternating inter-wire current and relative-phase structure have previously been interpreted as an array of Josephson vortices in Ref.~\cite{qiu2021stripe}. The state can be written as
\begin{equation}
\psi_{\rm V}(x)
=
\frac{1}{\sqrt{2}}
\left[
\psi_{k}(x)
+
e^{i\chi}\psi_{-k}(x)
\right],
\end{equation}
where $\chi$ is the relative phase between the amplitudes in the two
minima. In the thermodynamic limit, the continuous degeneracy associated with $\chi$ reflects the spontaneous breaking of magnetic-translation symmetry. In the spin-orbit-coupled formulation, this is the origin of the supersolid character of the stripe phase and of the additional gapless collective branch associated with translational-symmetry breaking \cite{li2013superstripes}.

In the vortex phase, the density is balanced between the wires,
$\rho_c=0$, but the interference between the two momentum
components produces density modulations,
\begin{equation}
\rho_{1,2}(x)
=
\frac{1}{2}
+
\frac{\cos(2kx-\chi)}{f_k+f_k^{-1}}, \label{eq:vortexdens}
\end{equation}
with wavevector $2k$; hence why this is known as the stripe phase. In the thermodynamic limit, the period of these modulations is
\begin{equation}
\lambda_{\rm stripe}
=
\frac{\pi}{|k|}
=
\frac{\ell_B}{\sqrt{1-4\tilde J^2}}.
\end{equation}
The currents along each wire are also spatially modulated, such that
the local chiral current $j_c(x)\equiv j_2(x)-j_1(x)$ is given by
\begin{equation}
j_c(x)
=
\frac{\phi}{2}
+
\frac{k(f_k-f_k^{-1})}{f_k+f_k^{-1}}
+
\frac{\phi}{f_k+f_k^{-1}}
\cos(2kx-\chi).
\end{equation}
This averages spatially to
\begin{equation}
\overline{j}_c
= \frac{1}{L}\int_0^L dx \  j_c(x) = 
\frac{\phi}{2}
+
\frac{k(f_k-f_k^{-1})}{f_k+f_k^{-1}}.
\end{equation}
Using
\begin{equation}
\frac{f_k-f_k^{-1}}{f_k+f_k^{-1}}
=
-\frac{k/\phi}{\sqrt{(k/\phi)^2+\tilde J^2}},
\end{equation}
and the thermodynamic minimum condition
\begin{equation}
(k/\phi)^2+\tilde J^2=\frac{1}{4},
\end{equation}
this becomes
\begin{equation}
\overline{j}_c
=
2\phi\tilde J^2.
\end{equation}
The inter-wire current also oscillates with the same spatial period as
the density modulation, while being shifted in phase relative to the
longitudinal-current pattern. This generates a sequence of circulating
current loops along the ladder, providing the characteristic vortex
pattern~\cite{qiu2021stripe}. As can be seen, the relative phase $\chi$ of the two momentum components
translates this entire pattern along the wire without changing its
period or spatially averaged properties. Note that in the Hofstadter ladder, the vortex phase is naturally described in terms of current loops circulating around discrete plaquettes, whereas here the notion of ``plaquettes" is emergent rather than microscopic. The vortex pattern is instead identified from the spatially periodic modulation of the density, longitudinal and inter-wire currents.

We next turn to the biased-density wire phase, which occurs when the
condensate wavefunction is simply either $\psi_k$ or $\psi_{-k}$. As
suggested by the choice of name, this phase is closely analogous to the
biased-density ladder state previously studied for the Harper-Hofstadter
model~\cite{Wei2014TwoLegBosons}, and it is known as the plane-wave phase for a spin-orbit-coupled condensate as only a single plane-wave is present~\cite{li2012tricriticality,qiu2021stripe}. Using Eq.~\ref{eq:BudichWavefunc}, it
can be shown that the densities along each wire for $\psi_k$ are given
by
\begin{equation}
\rho_1
=
\frac{f_k^{-1}}{f_k+f_k^{-1}},
\qquad
\rho_2
=
\frac{f_k}{f_k+f_k^{-1}},
\end{equation}
which are spatially uniform along $x$, but imbalanced between the wires
as $\rho_1\neq\rho_2$. The spatially averaged density imbalance, $\rho_c\equiv\overline{\rho}_1-\overline{\rho}_2$, equal to its constant value in the biased-density phase, is given by
%
%%
% \begin{equation}
% \rho_c
% \equiv
% \overline{\rho}_1-\overline{\rho}_2
% =
% \frac{1}{L}
% \int_0^L dx\,
% \left[
% \rho_1(x)-\rho_2(x)
% \right],
% \label{eq:density_imbalance_definition}
% \end{equation}
% %
% which 
\begin{equation}
\rho_c
=
\frac{f_k^{-1}-f_k}{f_k+f_k^{-1}}
=
\frac{k/\phi}{\sqrt{(k/\phi)^2+\tilde J^2}}.
\label{eq:rhoc}
\end{equation}
 Hence $
|\rho_c|
=
\sqrt{1-4\tilde J^2} = |2k_\pm/\phi| $.
At the transition from the vortex phase, when
$\tilde J=1/\sqrt{6}$, this jumps from zero to 
\begin{equation}
|\rho_c|=\frac{1}{\sqrt{3}}, \label{eq:densityjump}
\end{equation} 
before decreasing continuously to zero as $\tilde J\rightarrow1/2$, where the ground-state transitions into the Meissner-like phase. The two choices of $\psi_k$ and $\psi_{-k}$ have opposite density imbalances corresponding to $\rho_c >0$ and $\rho_c <0$ respectively. However, these two degenerate biased-density states still carry the same chiral current. This can be seen by noting that the
intra-wire currents are
\begin{equation}
j_1
=
\rho_1\left(k-\frac{\phi}{2}\right),
\qquad
j_2
=
\rho_2\left(k+\frac{\phi}{2}\right),
\label{eq:currents}
\end{equation}
where we have dropped the explicit dependence on $x$, as these currents are spatially uniform. 
Hence,
\begin{equation}
j_c
\equiv
j_2-j_1
=
\frac{\phi}{2}
-
k\rho_c .
\end{equation}
As the two choices of momentum have opposite signs of both $k$ and
$\rho_c$, their product, and hence the chiral current, is
identical for the two degenerate states. Using Eq.~\ref{eq:rhoc}, this
can be written as
\begin{equation}
j_c
=
\frac{\phi}{2}
-
\frac{\phi(k/\phi)^2}
{\sqrt{(k/\phi)^2+\tilde J^2}} =2\phi\tilde J^2. \label{eq:denscurr}
\end{equation}
Thus the vortex and biased-density phases have the same (spatially-averaged) chiral current within the same momentum sector. The transition
between them is therefore not diagnosed by a discontinuity in the
averaged chiral current, but instead by both the disappearance of the spatial
density and current modulations and the onset of a uniform population
imbalance between the wires.

Finally, for $\tilde J\geq1/2$, the lower band has a single minimum at
$k=0$. The wavefunction is then
\begin{equation}
\psi_{k=0}(x)
=
\frac{1}{\sqrt{2}}
\begin{pmatrix}
e^{-i\phi x/2}\\
e^{i\phi x/2}
\end{pmatrix}.
\end{equation}
It can immediately be seen that the densities on the two wires are equal and spatially uniform, while
the phases wind linearly in opposite directions. The net longitudinal
current vanishes, while the chiral current is finite and equal to
\begin{equation}
j_c=\frac{\phi}{2}.
\end{equation}
This ``saturated chiral current" is maximal for this model, connecting smoothly to the value of the chiral current at the transition from the biased-density phase [c.f. Eq.~\ref{eq:denscurr}]. This is the analogue of the Meissner-like phase found in the  Harper-Hofstadter ladder~\cite{Wei2014TwoLegBosons,budich2017coupled} and is known as the zero-momentum phase for a spin-orbit-coupled condensate~\cite{li2012tricriticality,qiu2021stripe}.

\paragraph{Finite-size effects.---} 
We now turn to a finite-size system, which has some subtleties that can be relevant for future experiments. A significant complication in this case is that the momenta of the true
thermodynamic minima, $k_\pm$, are not generically compatible with the
periodic boundary conditions. Specifically, we must require that both
components in Eq.~\ref{eq:BudichWavefunc} are single-valued, such that
\begin{equation}
e^{i(k-\phi/2)L}=1,
\qquad
e^{i(k+\phi/2)L}=1 .
\end{equation}
Hence, we can write
\begin{equation}
(k-\phi/2)L=2\pi n_-,
\qquad
(k+\phi/2)L=2\pi n_+,
\end{equation}
where $n_\pm\in\mathbb{Z}$. Adding and subtracting these leads to two
conditions
\begin{equation}
k=\frac{\pi(n_++n_-)}{L},
\qquad
\phi=\frac{2\pi(n_+-n_-)}{L}.
\end{equation}
The former is a variant of the usual finite-size quantisation condition,
while the latter is simply a re-statement that the wire length must be
an integer multiple of the magnetic length, i.e.
\begin{equation}
L=z\ell_B=\frac{2\pi z}{\phi},
\end{equation}
with $n_+-n_-=z$. The allowed momenta may therefore be written as
\begin{equation}
k=\frac{n\phi}{2z},
\end{equation}
where $n\equiv n_++n_-$. Note that, as these are all integers, it
follows that $n$ always has the same parity as $z$. Thus for even $z$,
we can write $z\equiv2\tilde z$ and $n\equiv2\tilde n$, such that
$k=\tilde n\phi/2\tilde z$. Instead, for odd $z$, we write
$z\equiv2\tilde z+1$ and $n\equiv2\tilde n+1$, such that
$k=(2\tilde n+1)\phi/2(2\tilde z+1)$. Note that this implies that
$k=0$ is not an allowed momentum for any finite-size system with a
length along the continuous direction that is an odd integer multiple of
the magnetic length, which will prove to be important.

This even-odd finite-size effect can also be expressed in a more manifestly gauge-invariant form. For two wires, we may choose a centred gauge in which the effective magnetic vector potentials are $A_1=-\phi/2$ and $A_2=+\phi/2$ on the two wires respectively. The corresponding holonomy around the periodic continuous direction is
\begin{equation}
W_{1,2} = \exp\left(i\oint A_{1,2}dx\right) = \exp\left(\pm i\frac{\phi L}{2}\right).
\end{equation}
Using $L=z\ell_B=2\pi z/\phi$, this becomes
\begin{equation}
W_{1,2}=\exp(\pm i\pi z)=(-1)^z.
\end{equation}
Although a local gauge transformation can remove the vector potential from the Hamiltonian,  it will also change the boundary conditions of the transformed state as
\begin{equation}
\widetilde{\psi}_\lambda(x+L)
=
(-1)^z\widetilde{\psi}_\lambda(x).
\end{equation}
The transformed wavefunctions therefore obey periodic boundary conditions when $z$ is even and antiperiodic boundary conditions when $z$ is odd. Consequently, the allowed canonical momenta are
\begin{equation}
k=
\begin{cases}
\dfrac{2\pi m}{L}, & z\ \text{even}, \\[6pt]
\dfrac{2\pi}{L}\left(m+\dfrac{1}{2}\right), & z\ \text{odd},
\end{cases}
\qquad m\in\mathbb{Z}.
\end{equation}
This is equivalent to the above quantisation condition $k\!=\!n\phi/(2z)$, with $n$ having the same parity as $z$. 

When $\tilde J=0$, the spectrum has two minima which occur at
$k_\pm=\pm\phi/2$, corresponding to the case of uncoupled wires
[c.f. Fig.~\ref{fig:open-2-wire-Var}(a)]. Comparing this with the above
quantisation condition, this can be satisfied by $n=\pm z$, meaning
that the lowest-energy states are the same for the finite and
infinite-size system. As $\tilde J$ begins to increase, the momenta of
the true minima, $k_\pm$, smoothly decreases, while the lowest-energy
states of the finite-size system will remain fixed at $k=\pm\phi/2$ due to
momentum quantisation. Increasing $\tilde J$ further, the
lowest-energy states will switch between pairs of allowed momenta, i.e.
from $k=\pm\phi/2$ to $k=\pm(z-2)\phi/2z$ and so on. From
Eq.~\ref{eq:spectrum2wireOBE}, it can be shown that there is a finite-size
level crossing between states with $n$ and $n-2$ occurring whenever
\begin{equation}
\tilde J_{n\rightarrow n-2}
=
\frac{1}{2}
\sqrt{
\left[
1-\left(\frac{n-1}{z}\right)^2
\right]
\left[
1-\frac{1}{z^2}
\right]
}.
\label{eq:transition}
\end{equation}
This equation determines the crossings between different momentum sectors at the single-particle level. 

In our variational approach, we use this single-particle
spectrum to select the lowest-energy allowed momentum sector for a given
value of $\tilde J$. We then construct the interacting ansatz within
this restricted two-dimensional subspace, spanned by
$\psi_{\phi n/2z}$ and $\psi_{-\phi n/2z}$, and minimise the interaction
energy with respect to their relative occupations. This gives a low-energy variational description of the interacting ground state, while neglecting the possibility that interactions shift the boundaries between different momentum sectors by favouring a slightly higher single-particle state with lower interaction energy; this approximation is checked by the imaginary-time evolution results of the next section. 

When $n=0$, the ansatz trivially minimises the interaction energy to
give $\psi
=
\psi_{k=0}$, corresponding to the Meissner-like phase already discussed above. 
An example of the corresponding particle density and phase distributions
is shown in Fig.~\ref{fig:open-2-wire-Var}(h) for $J=2.5$,
$\phi=\pi/2$ and $L=2\ell_B$. When $n\neq0$, our variational ansatz becomes
\begin{equation}
\psi
=
a_1\psi_{\phi n/2z}
+
a_2\psi_{-\phi n/2z},
\qquad
|a_1|^2+|a_2|^2=1,
\end{equation}
where we have taken $n>0$ for simplicity of notation. Note that, for a
given $\tilde J$, the appropriate value of $n$ must be chosen such that the
corresponding states are the lowest-energy states available in the
finite-size single-particle spectrum [c.f. Eq.~\ref{eq:transition}]. As in the thermodynamic limit, we then minimize the interaction energy to get the polynomial in Eq.~\ref{eq:EintThermo}, except now with $k=\phi n / 2z$. This interaction energy is again minimised by by either
$|a_1|^2=1/2$ or $|a_1|^2=0,1$, corresponding to the vortex-like and biased-density phase, respectively. However, now the boundary between the two
phases occurs at
\begin{equation}
\tilde J_c^{(n)}
= \sqrt{2}\frac{k}{\phi}= 
\frac{n}{z\sqrt{2}},
\label{eq:vortex-biased-transition}
\end{equation}
which depends on $n$ and $z$ in general. In the thermodynamic limit, the allowed momenta
become dense, and the sector selected by the
single-particle spectrum approaches the continuum band minimum and $\tilde J_c \rightarrow 1/\sqrt{6}$, as stated above. Numerically, we observe the transition between these states by examining the density imbalance $\rho_c\equiv \rho_1-\rho_2$ as plotted in Fig.~\ref{fig:open-2-wire-Var}(b) as a function of magnetic length $l_B=2\pi / \phi$ and hopping $J$, for $g=1$ and $L=\ell_B$. Here our analytical prediction [Eq.~\ref{eq:vortex-biased-transition}] is in excellent agreement with the sharp onset of a large density difference between the two wires, indicating the transition between the two phases. Note that for odd $z$, like here, the $k=0$ state is inaccessible,
meaning that as $\tilde J$ increases, the system never transitions to
the exact Meissner-like phase. Instead, the large-$\tilde J$ state
remains in the final allowed biased-density sector, although its
properties continuously approach those of the Meissner phase in the
thermodynamic limit. 

Interestingly, whether a biased-density phase is realized in a given finite system in the weakly-interacting limit depends
on the competition between the interaction-driven transition within a
given momentum sector and the single-particle switching between
different momentum sectors. A sector $n$ supports an observable
vortex-to-biased transition only if $\tilde J_c^{(n)}$ lies within the
range of parameters over which that sector is selected by the
single-particle spectrum. Otherwise, the system switches to another
momentum sector before the biased-density state is reached, or enters
the sector only after the biased-density configuration is already
favoured.

For even $z$, the last transition occurs from $n=2$ to $n=0$, i.e. from
the pair of states at $k=\pm\phi/z$ to the single state $k=0$, when
\begin{equation}
\tilde J_{2\rightarrow0}
=
\frac{1}{2}
\left(
1-\frac{1}{z^2}
\right), \label{eq:20}
\end{equation}
which recaptures the expected result, $\tilde J=1/2$, as
$z\rightarrow\infty$. For odd $z$, the last possible transition is from
$n=3$ to $n=1$, i.e. from the pair of states at $k=\pm3\phi/2z$ to the
pair of states $k=\pm\phi/2z$, which only tends towards $k=0$ in the
thermodynamic limit.

This distinction leads to different finite-size sequences of states. On the one
hand, for odd $z$, the Meissner-like phase is inaccessible, as mentioned above, and so  the ground state will always be a biased-density phase for high enough $\tilde{J}$.
On the other hand, for $z=2$, the lowest energy
states appear either at $k=\pm\phi/2$, or $k=0$; it can then be found
that $\tilde J_c^{(2)}=1/\sqrt{2}$ while
$\tilde J_{2\rightarrow0}=3/8$. Hence, within this
single-particle-sector variational approximation, the system transitions
directly from the vortex-like phase to the Meissner-like phase, without
passing through the biased-density phase. By contrast, for $z=4$, the
allowed positive momenta are $k=\{\phi/2,\phi/4,0\}$. At the
single-particle level, the relevant momentum sector first switches from
$n=4$ to $n=2$ at $
\tilde J_{4\rightarrow2}
=
{\sqrt{105}}/{32}$,
followed by the transition to a biased-density phase within the $n=2$
sector at $
\tilde J_c^{(2)}
=
{1}/{(2\sqrt{2})}$,
and finally the transition to the $k=0$ Meissner-like phase at
$
\tilde J_{2\rightarrow0}
=
{15}/{32}$.

Finite-size effects will also modify the properties of the vortex-like phase and the density-biased phase. Firstly, in the vortex phase, the density modulation is given by Eq.~\ref{eq:vortexdens}, but with $k=\phi n / 2 z$; hence the stripe period becomes
\begin{equation}
\lambda_{\rm stripe}
=
\frac{\pi}{|k|}
=
\frac{z\ell_B}{|n|},
\end{equation}
corresponding to having $|n|$ periods over the length of the wire. Similar behaviour is observed for the longitudinal currents. Examples of the corresponding density and phase profiles in the vortex-like phase are shown in Fig.~\ref{fig:open-2-wire-Var} for $J=0.2$,
$\phi=\pi/2$ with (e) $L=\ell_B$ and (g) $L=2\ell_B$.  The
spatially averaged chiral current is also affected by the finite-size effects, becoming
\begin{equation}
\overline{j}_c^{(n)}
=
\frac{\phi}{2}
-
\frac{\phi(n/2z)^2}
{\sqrt{(n/2z)^2+\tilde J^2}} . \label{eq:finitejc}
\end{equation}
This can therefore change discontinuously when the lowest
single-particle sector switches between different allowed values of
$n$.

Secondly, in the biased-density wire phase, the density imbalance of Eq.~\ref{eq:rhoc} now becomes,
\begin{equation}
\rho_c^{(n)}
=
\frac{n/2z}
{\sqrt{(n/2z)^2+\tilde J^2}}.
\end{equation}
Thus, at fixed $n$, the density imbalance decreases as $\tilde J$ is
increased. Similar to the thermodynamic case, this two-mode variational theory predicts that, within a given momentum sector, the density difference jumps discontinuously at the transition from the vortex phase into the biased-density phase with $|\rho_c| = 1/\sqrt{3}$, which is independent of the system size [c.f. Fig.~\ref{fig:open-2-wire-Var}(b)]. However, the transition to the biased-density phase does not always occur in the same momentum sector. If $z=3$, then $\tilde{J}_c^{(1)} < \tilde{J}_{3\rightarrow1} < \tilde{J}^{(3)}$, meaning the system will immediately change into the biased-density phase at the level crossing between states $\tilde{J}_{3\rightarrow1}$ resulting in a different $\rho_c$ jump at the onset of the biased-density phase. The density imbalance can also change discontinuously at further transitions between different allowed momentum sectors as well as at the transition into the
density-balanced Meissner-like phase. For example, for even $z$, if the system is in the biased-density state with $n=2$, then immediately before the transition to the Meissner-like phase the density difference is
\begin{equation*}
\rho_c^{(2)}
=
\frac{2z}{z^2+1},
\end{equation*}
which only vanishes in the thermodynamic limit. For the finite-size biased-density phase, the chiral current is again uniform and equal to the spatially-averaged chiral current in the vortex phase [c.f. Eq.~\ref{eq:finitejc}]. Interestingly, in a finite-size system. the biased-density state can also carry a
non-zero total longitudinal current,
\begin{equation}
j_{\rm tot}
\equiv
j_1+j_2
=
k-\frac{\phi}{2}\rho_c
=
k\left[
1-
\frac{1}{2\sqrt{(k/\phi)^2+\tilde J^2}}
\right].
\end{equation}
This current vanishes when the occupied momentum coincides with the
true minimum of the continuous single-particle band,
$(k/\phi)^2+\tilde J^2=1/4$, and is therefore a finite-size effect
associated with momentum quantization. It is generally small when the
allowed momentum lies close to the continuum minimum, but can be
appreciable for very short odd systems, such as $L=\ell_B$, where the
$k=0$ Meissner state is inaccessible. An example of the corresponding
density and phase distributions in this biased-density phase is shown
in Fig.~\ref{fig:open-2-wire-Var}(f) for $J=2.5$, $\phi=\pi/2$ and
$L=\ell_B$. Note that for this system size and hopping amplitude, the quantized minima sit at $k=\pm \phi/2$. The intra-wire currents [Eq.~\ref{eq:currents}] are consequently given by $j_2=0$ and $j_1=\rho_2 \phi$, corresponding to having a phase which is constant along one wire and a phase which winds by $2\pi$ across the system along the other wire, as can be seen in the figure.

\begin{figure*}
    \centering
    {\includegraphics[trim={0.0cm 0.0cm 0.0cm 0.0cm},clip,width=\linewidth]{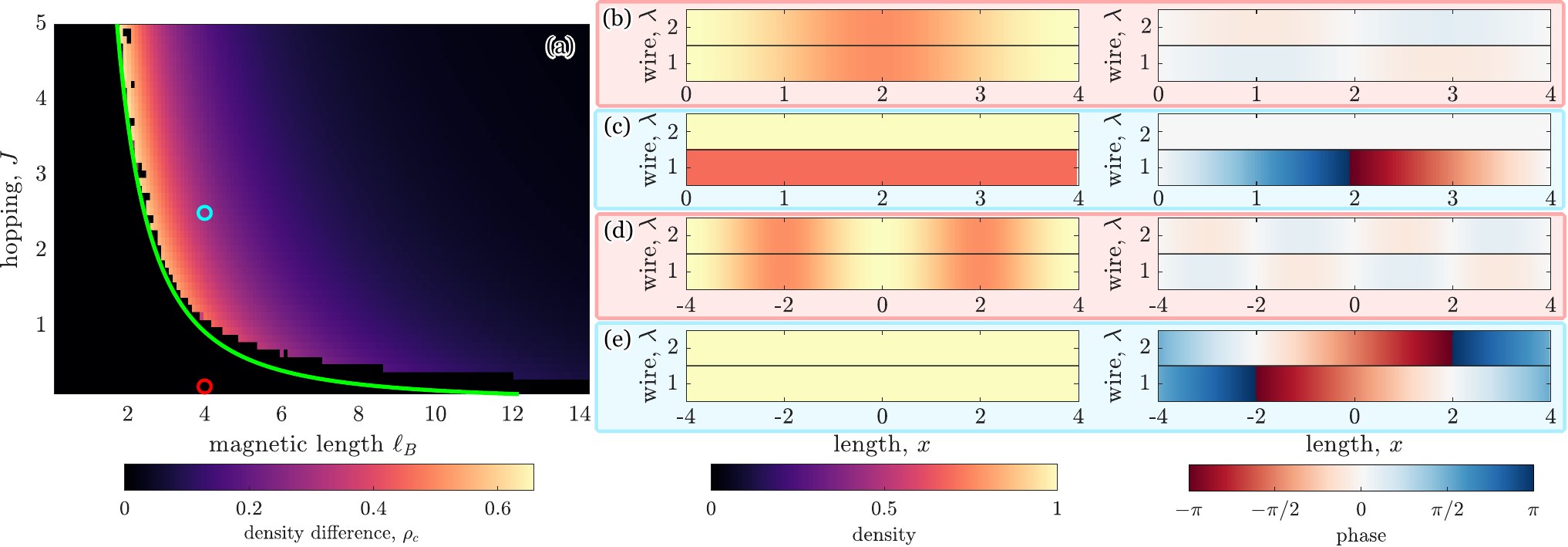}}
    \caption{Interacting ground-states of two coupled wires with open boundary conditions along the discrete direction as obtained
numerically from numerically solving the Gross-Pitaevskii equation using imaginary-time evolution. (a) The density difference between the two wires, $|\rho_c|$, as a function of the inter-wire hopping $J$, and the magnetic length of the system, $\ell_B$, for $L=\ell_B$. The analytical prediction [Eq.~\ref{eq:vortex-biased-transition}] for the transition between the vortex-like and density-biased phase is indicated by the green line. The red and blue dots mark the parameters $\phi=\pi/2$, with $J=0.2$ and $J=2.5$ respectively. Panels (b) and (c) show the density and phase profiles for the red and
blue points in panel (a), respectively, with $L=\ell_B$. These correspond
to the vortex-like and biased-density phases. Panels (d) and (e) show the
same parameters but with $L=2\ell_B$, giving the vortex-like and
Meissner-like phases, respectively. }
    \label{fig:open-2-wire-GPE}
\end{figure*}

\subsubsection{Imaginary time evolution}

As discussed above, the variational method uses a limited single-particle basis to minimise the interaction energy. To check these results, we also use imaginary-time evolution to minimise the full Gross-Pitaevskii energy functional [Eq. \ref{eq:gpe_1}] without restricting the condensate to the two-mode variational subspace. This method is more computationally expensive than the variational method and does not give closed-form analytical results. However, it minimises
the full Gross-Pitaevskii energy functional without restricting the
condensate to the two-mode variational subspace, and can therefore better capture
the profile of the condensate profile, going beyond the variational ansatz.

In Fig. \ref{fig:open-2-wire-GPE}(a) we present the density difference of the two wires as a function of the inter-wire hopping, $J$, and the magnetic length of the system, $\ell_B$, with the above analytical prediction [Eq. \ref{eq:vortex-biased-transition}] for the transition between the vortex-like and density-biased phase plotted in green. Despite the differences between the two methods, we still see a strong agreement between this analytical prediction and the sharp onset of the density difference between the legs, indicating a change in the ground state. Note that there is an increasing divergence in these approaches as the inter-wire hopping $J$ decreases and  magnetic length increases. Here the separation between the low-energy
single-particle states and the higher modes becomes small; in this regime,
the interaction energy can more easily admix states outside the restricted
two-mode variational subspace, meaning that the assumptions underlying the
variational ansatz become less accurate.
Away from this limit, however, the full Gross-Pitaevskii equation is able to capture all of the predicted physics given by the variational method, suggesting that the different phases should be easily accessible in a coupled 1D Bose-Einstein condensate setup.

\begin{figure}
    \centering
{{\includegraphics[trim={0.0cm 0.0cm 0cm 0.0cm},clip,width=0.95\linewidth]{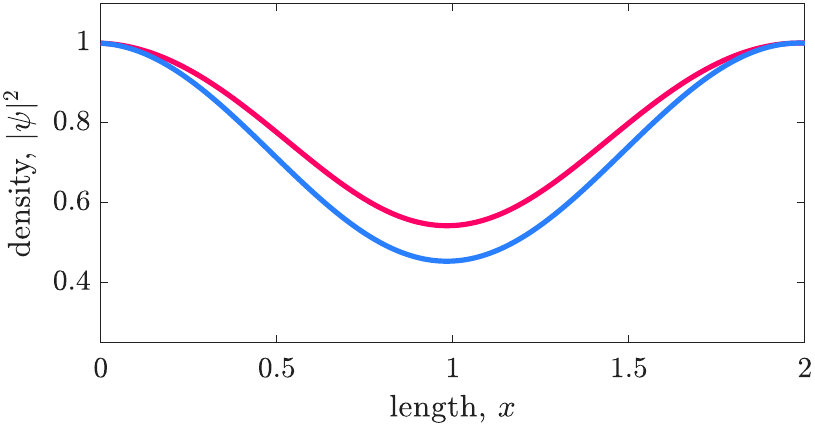}}}
    \caption{The ground state density along one wire for the vortex-like state found for $J = 1$, $L=\ell_B$ and $\ell_B = 2$ via the variational method (\textit{blue}) [c.f. Fig.~\ref{fig:open-2-wire-Var}] and imaginary time evolution method (\textit{red}) [c.f. Fig.~\ref{fig:open-2-wire-GPE}]. Both curves exhibit density modulations with the same spatial period, but with slightly varying shapes and amplitudes.}
    \label{fig:vortex difference}
\end{figure}

As in the variational case, we can also look at the density and phase profiles of the calculated ground states. We can immediately see that the ground state in both the $L=\ell_B$ (Fig. \ref{fig:open-2-wire-GPE}(b,c)) and $L=2\ell_B$ cases (Fig. \ref{fig:open-2-wire-GPE}(d,e)) agrees with those predicted previously in Section \ref{sec:variational_2_wire}; the vortex case again consists of a single vortex between the two wires over every $\ell_B$. The density-biased and Meissner-like phases again have a constant density over the wire length, with a density difference between the wires which is finite for the former and zero for the latter. Importantly, we note that the imaginary-time results also reproduce the finite-size
parity effect predicted by the variational method; for the same large
hopping amplitude, the system remains in a biased-density state when
$L=\ell_B$, where the $k=0$ state is inaccessible, but reaches the
Meissner-like state when $L=2\ell_B$, where $k=0$ is an allowed
finite-size momentum.

\begin{figure}
    \centering
{\includegraphics[trim={0.0cm 0.0cm 0cm 0.0cm},clip,width=\linewidth]{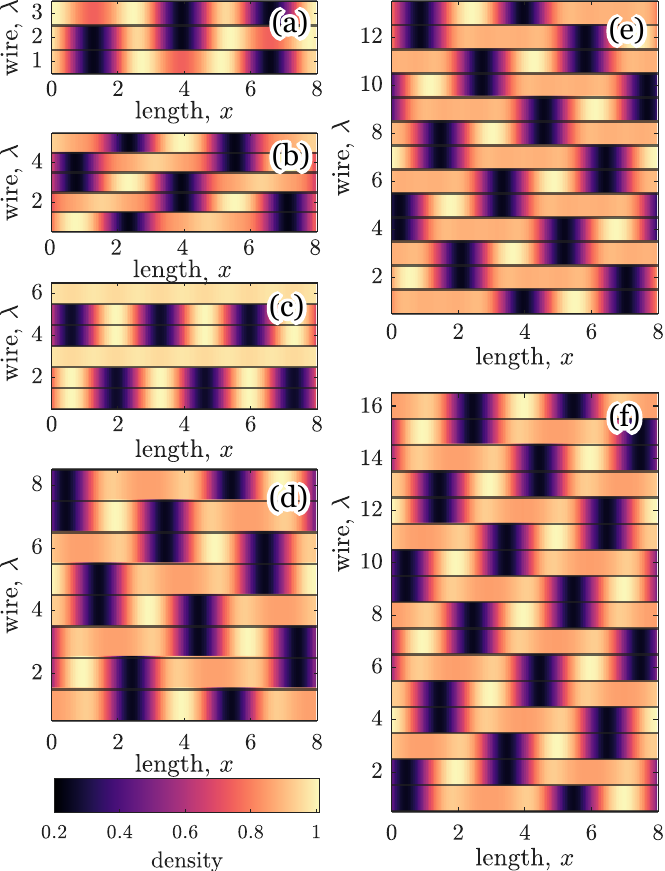}}
    \caption{The density of the ground state for a coupled wire system with periodic boundary conditions along the discrete direction, solved via the imaginary-time evolution method for $\phi=\pi/4$, $L=\ell_B$ and $J=1$ with (a) $N=3$, (b) $N=5$, (c) $N=6$, (d) $N=8$, (e) $N=13$ and (f) $N=16$ wires.}
    \label{fig:large_periodic}
\end{figure}

Finally, we note that an observable difference between the variational and the imaginary time methods is the profile of the density, as shown in Fig \ref{fig:vortex difference}. This discrepancy arises because the variational method
projects the condensate onto a restricted two-state subspace, whereas the full Gross-Pitaevskii equation allows the nonlinearity to admix higher single-particle modes and deform the density profile. This accounts for the departure from the sinusoidal modulation predicted by the two-mode ansatz [c.f. Eq.~\ref{eq:vortexdens}]. We note that this nonlinear deformation is also captured by the closed-form Josephson-vortex solutions of Ref.~\cite{qiu2021stripe}.

\section{Systems with large numbers of wires}
\label{sec:large}

In this section, we focus only on the vortex regime, and look at how this changes as one increases the number of wires, $N$, thus building our hybrid continuous-discrete analogue of a two-leg ladder into a fuller two-dimensional geometry.  In two wires, the vortex state is strongly constrained by the small number of discrete sites, being characterised by a distinctive density modulation and current pattern for the two wires, as discussed above. We now explore how this few-wire vortex structure develops into a vortex lattice, analogous to those found in rotating condensates~\cite{abo2001observation,coddington2003observation} and Hofstadter lattice systems~\cite{powell2011bogoliubov, reid2026phases}. This question is closely related to previous work on synthetic-dimension quantum Hall systems, where a large synthetic dimension (with $N=17$) formed from the internal states of dysprosium atoms was predicted to support mean-field ground states resembling Abrikosov-like vortex lattices, with different confined geometries distinguished by the number of vortex lines along the real spatial direction~\cite{chalopin2020probing}. In this section, we explore the behaviour of our simple coupled wire model over a range of different $N$, again looking at the two cases of having either periodic or open boundary conditions in the discrete direction. 
As $N$ increases, the imaginary time method becomes very sensitive, and can occasionally fail to find the ground state solution. To keep the computations tractable, we keep the hopping fixed to $J=1$.

To understand the following large-$N$ results, it can be helpful to start from a general flux-counting argument. As $\ell_B = 2 \pi / \phi$, each strip of length $\ell_B$ between neighbouring wires encloses one flux quantum, corresponding to a single magnetic unit cell~\cite{budich2017coupled}. Increasing the number of wires increases the number of such magnetic unit cells, and hence the total number of flux quanta, in a given system. For example, for a system of length $L=\ell_B$, there will be a total of $N$ [resp. $N-1$] magnetic unit cells for periodic [resp. open] boundary conditions in the discrete direction. The total number of expected vortices can therefore be expected to scale with the number of wires in the system. The boundary conditions in the discrete direction then play a key role in determining how these vortices arrange themselves: we find numerically that periodic boundaries favour an Abrikosov-like vortex lattice, while open boundaries appear to partially reconstruct this lattice into vortex rows or lines [c.f.~Ref.~\cite{chalopin2020probing}].

\subsection{Periodic discrete boundary conditions} \label{sec:periodic_large}

In the two-wire periodic case discussed above, the ground state resembles a vortex-anti-vortex pair over each magnetic length of the system, for all parameters of inter-wire hopping and magnetic flux. This could be in part understood as emerging from the single-particle eigenstates, which take the form of Mathieu functions. 
For arbitrary $N$, some of this analytic structure remains; if we Fourier transform along the discrete direction,
\begin{equation}
\psi_\lambda(x)
=
\frac{1}{\sqrt{N}}
\sum_{m=0}^{N-1}
e^{ik_m\lambda}\chi_m(x),
\qquad
k_m=\frac{2\pi m}{N},
\end{equation}
the non-interacting problem separates into different transverse-momentum sectors,
\begin{equation}
\left[
-\frac{1}{2}\frac{d^2}{dx^2}
-
2J\cos(\phi x+k_m)
\right]\chi_m(x)
=
\epsilon_m\chi_m(x).  \label{eq:ma}
\end{equation}
Within each sector, we therefore again have a Mathieu problem with the same modulation period as before, but now with a shift of the potential along the wire by $k_m/\phi=m\ell_B/N$. When $L=z\ell_B$, each of these shifted equations have the same spectrum. Thus, for periodic discrete boundaries, the low-energy single-particle manifold consists of $N$ degenerate orbitals, shifted relative to one another by $\ell_B/N$ along the continuous direction. This is the coupled-wire analogue of the guiding-centre degeneracy in a Landau level. Equation~\eqref{eq:ma} is also directly analogous to the Harper equation obtained from the Hofstadter model in Landau gauge~\cite{harper1955UniformBfield}. In the
Harper problem, the conserved momentum shifts the phase of the effective 1D potential; here, the transverse momentum $k_m$ plays the same role, while the discrete coordinate of the Harper equation is replaced by the continuous wire coordinate $x$.

As in the Harper or lowest-Landau-level problem, an individual guiding-centre orbital does not by itself form a vortex lattice. Vortex states instead arise from coherent superpositions of the translated orbitals, as captured directly by our variational ansatz. Minimizing the interaction energy selects the relative amplitudes and phases within this superposition to determine the interacting ground state. Spreading the density minima between neighbouring wires reduces the overlap of depleted regions, naturally favouring a staggered configuration. In the large-$N$ limit this is the same qualitative mechanism underlying triangular Abrikosov order, although the Mathieu-orbital structure means that the present lattice is not identical to the continuum lowest-Landau-level solution.
Examples of the corresponding density patterns are shown for increasing values of $N$ in Fig.~\ref{fig:large_periodic} for $\phi=\pi/4$ and $J=1$, with $g=1$ and $L=\ell_B$. These results have been obtained using imaginary time evolution to solve the full Gross-Pitaevskii equation, although the same qualitative ground-state structure is found via the variational method, up to deviations in the density profile already discussed above [c.f. Fig.~\ref{fig:vortex difference}].

Although Fig.~\ref{fig:large_periodic} shows only the density patterns, the identification of the density minima as vortices can be confirmed by the corresponding phase and current patterns. In particular, the gauge-covariant circulation has the same sign around neighbouring density minima, in contrast to the vortex-antivortex current pattern found in the two-wire periodic case. This shows that, once the system has enough transverse extent, the periodic geometry supports a lattice of like-signed vortices, similar to in 2D rotating condensates~\cite{abo2001observation} or 2D Hofstadter models~\cite{powell2011bogoliubov, reid2026phases}. The two-wire case is therefore a special minimal geometry, in which the current pattern is too constrained to realise a conventional array.

As can be seen from Fig.~\ref{fig:large_periodic}, the observed number of vortices appears to be equal to the number of wires, in agreement with the above simple flux-counting argument. For large enough systems (e.g. $N \geq 8$), these vortices are staggered along the discrete direction as one moves along the continuous direction, resembling a triangular Abrikosov lattice~\cite{abrikosov2004nobel,abo2001observation,coddington2003observation,cooper2005vortex,cooper2008rapidly}. However, for small systems [c.f. panel (c)], this staggering may not occur due to the interplay of the finite transverse circumference, the fixed magnetic period $\ell_B$, and the integer number of vortices. In such cases, the interaction-selected superposition of the shifted Mathieu orbitals can be frustrated by the small number of wires, leading to a less triangular arrangement.

%Extending this to systems with larger number of wires, we observe that an Abrikosov lattice always forms. The number of vortices formed is always equal to the number of wires, with the orienatation of the vortices changing cyclically as a function of wire number, always forming a periodic triangular lattice. The results presented in Fig. \ref{fig:large_periodic} are given from the Gross-Pitaevskii equation, however the same result can be obtained from the variational method, with the previously explained caveat that the shape of the individual vortices is different.
%Extending this to larger systems, we see that no matter the number of wires, that there will always be one vortex formed between each neighbouring wire, and the geometry of the lattice formed will change to accommodate the number of wires.  

\subsection{Open discrete boundary conditions}

We now turn to the case of open boundary conditions along the discrete direction. Before proceeding to the numerical results, it is useful to recall the single-particle spectrum shown in Fig.~\ref{fig:band_structure}. As noted previously, when $J=0$, the spectrum consists of $N$ decoupled parabolas whose minima are separated in momentum by $\phi$. Although finite hopping hybridises neighbouring parabolas, the low-energy structure still resembles approximately equally spaced minima [c.f. Fig.~\ref{fig:band_structure}(a)].  A vortex state can therefore be viewed, at least qualitatively, as an interacting condensate formed in a coherent superposition of several of these low-energy states.

\begin{figure}[!]
    \centering
{\includegraphics[trim={0.0cm 0.0cm 0cm 0.0cm},clip,width=\linewidth]{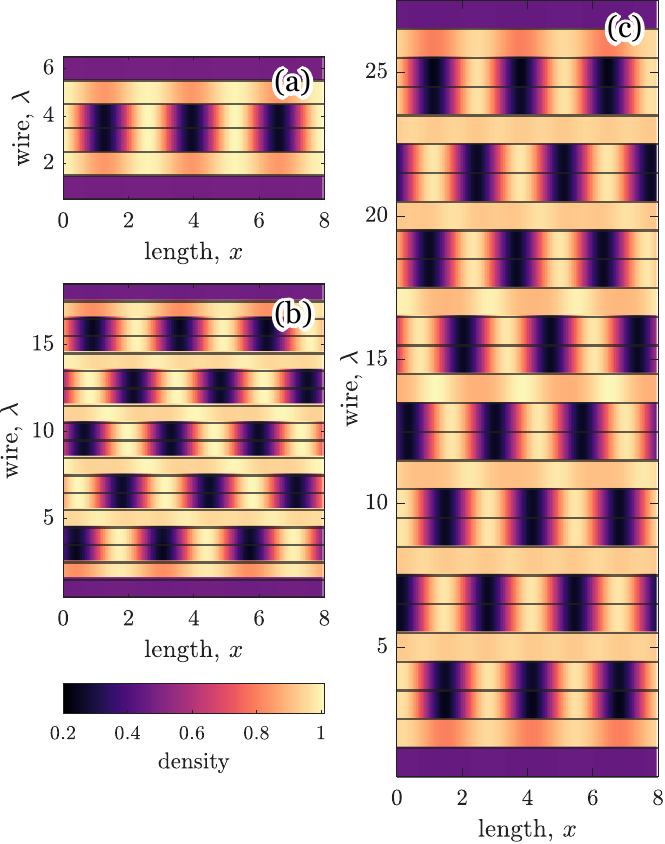}}
    \caption{The density of the ground state for a coupled wire system with open boundary conditions along the discrete direction solved via the imaginary time method for $\phi=\pi/4$, $L=\ell_B$ and $J=1$ with (a) $N=6$, (b) $N=18$ and (c) $N=27$.}
    \label{fig:large_open}
\end{figure}

The longitudinal period of the resulting density and current patterns is set by the magnetic translation symmetry. The hopping phases are periodic under $x\to x+\ell_B$, meaning that the ground-state density is naturally expected to repeat on this scale. The single-particle spectrum gives the same intuition: interference between states in neighbouring low-energy minima, whose momenta differ by approximately $\phi$, produces a leading-order density modulation with period $2\pi/\phi=\ell_B$.

In principle, one can use these low-energy minima to construct an $N$-state variational ansatz for the interacting ground state. In practice, however, this becomes less systematic for $N >2$. In the two-wire case, the ansatz can always be built from a small number of degenerate lowest-energy states. For larger $N$, however, the degeneracy between the minima is lifted for non-zero $\tilde{J}$ [c.f.~Fig.~\ref{fig:band_structure}], and there is no clear prescription for which states should be retained. Heuristically, we found that excluding the outermost energy minima of the band structure (i.e. the two minima with the largest values of $|k_x|$) was sufficient to reproduce our imaginary time-evolution results for $5 \leq N < 8$, up to differences like those in the vortex shape discussed above. However, to avoid this arbitrariness, we only present the imaginary-time results in what follows.

The resulting density profiles are shown in Fig.~\ref{fig:large_open} for $\phi=\pi/4$, $J=1$, $g=1$, $L=\ell_B$ and increasing values of $N$. As in the periodic case, the identification of these density minima as vortices is confirmed by the corresponding phase and current patterns. For small numbers of wires, we can see that the vortices are strongly affected by the finite transverse width. For example, in the $N=6$ case, the density minima are concentrated along a single row, corresponding to having a line of vortices between the same neighbouring wires. As $N$ is increased, additional vortex rows can fit into the system, with relative offsets that are reminiscent of an Abrikosov lattice. Similar vortex-line structures have also been predicted for synthetic-dimension quantum Hall systems, where the finite number of internal states ($N=17$) naturally imposes open boundary conditions in the synthetic direction~\cite{chalopin2020probing}.

To understand the appearance of vortex rows, it is useful to consider the transverse hopping problem at fixed $x$. For open boundary conditions, the position-dependent Peierls phase can be removed from the transverse hopping matrix by an $x$-dependent change of phase, without changing the density on each wire. The corresponding transverse eigenmodes therefore have the usual open-chain form
\begin{equation}
u_{\lambda}^{(r)}
=\sqrt{\frac{2}{N+1}} \sin\left(\frac{r\pi\lambda}{N+1}\right),\qquad
r=1,\ldots,N.
\end{equation}
In particular, the lowest transverse mode has the density envelope
\begin{equation}
\left|u_{\lambda}^{(1)}\right|^2 =
\frac{2}{N+1} \sin^2\left(\frac{\pi\lambda}{N+1}\right),
\end{equation}
which is largest near the centre of the discrete direction and is suppressed towards the open boundaries. Although the longitudinal kinetic energy and synthetic gauge field modify the detailed condensate profile, this simple result provides an analytical explanation for the central density envelope observed in Fig.~\ref{fig:large_open}. This is consistent with the numerical observation that well-defined vortex cores occur predominantly near the centre of the discrete direction, while near the edges, part of the imposed gauge field can instead be accommodated by boundary currents and smooth phase gradients.

This argument also explains the observed even-odd dependence. For even $N$, the maximum of the envelope is shared by the two central wires, so that a vortex row can lie in the central inter-wire region. For odd $N$, the maximum instead lies on the central wire. Reflection symmetry then favours vortex rows appearing in pairs on either side of the centre. The simple flux-counting argument therefore gives the overall scale of the vortex number, but can overestimate the number of well-defined bulk vortices in a system with open boundaries along the discrete direction.

\section{Conclusions}
We have studied how synthetic gauge fields generate vortex-like ground states in arrays of coupled one-dimensional Bose-Einstein condensates, forming a hybrid system with one continuous and one discrete spatial direction. The resulting physics depends strongly on both the number of wires and the boundary conditions imposed along the discrete direction.

For two wires with open discrete boundaries, both variational and imaginary-time-evolution methods reproduced vortex-like, Meissner-like and biased-density phases known from spin-orbit coupled gases and Harper-Hofstadter ladders. In the thermodynamic limit, the weakly-interacting ground state of the two coupled wires first undergoes a transition from the vortex-like to the biased-density phase, followed by a transition to the Meissner-like state. For finite wires, the periodicity along the continuous direction quantises the allowed condensate momenta, producing a parity effect for $z=L/\ell_B$. When $z$ is even, the $k=0$ Meissner-like state is accessible, whereas for odd $z$ it is excluded and the large-hopping ground state remains in a finite-momentum biased-density sector.  

With periodic boundaries along the discrete direction, we showed that the two-wire system instead supports a minimal pair of counter-circulating current cells, similar to a vortex-anti-vortex pair, within each magnetic length. As the number of wires increases, this develops into an extended array of like-signed vortices. For sufficiently large $N$, the vortex positions are staggered along the discrete direction, resembling an Abrikosov-like triangular arrangement. The number of vortices scales with the number of magnetic unit cells, in accordance with a simple flux-counting argument.

Open boundaries along the discrete direction qualitatively reconstructs the vortex lattice in the coupled wire model. In this case, the transverse density is suppressed near the edges and concentrated towards the centre of the system, such that well-defined vortices first form in the central rows of the system. As $N$ increases, additional rows appear, gradually approaching the staggered structure found with periodic boundaries. With open boundaries, there is also an even-odd dependence on the number of wires and a reduction in the expected number of visible bulk vortices due to the decreased density near the system edges. For the systems with a large number of wires with both periodic and open boundaries along the discrete direction, the hopping parameter $J$ was fixed. An interesting extension to this work would be to look at how $J$ affects the shape and structure of the vortex lattice formed.

The coupled-wire model that we have considered here is closely connected to existing proposals for arrays of real-space atomic wires. In particular, Ref.~\cite{budich2017coupled} proposed creating sub-wavelength separated wires using optical potentials, with Raman-assisted tunnelling between neighbouring wires to generate the position-dependent hopping phases in Eq.~\ref{eq:Landau}. In physical units, the dimensionless hopping used throughout this work can be written as $\widetilde{J} =J/{E_R^M}$ where $E_R^M =\hbar^2\phi^2/2m$ is the magnetic recoil energy. The experimentally-accessible regime of $\widetilde{J}\sim\mathcal{O}(1)$ discussed in Ref.~\cite{budich2017coupled} then may include results studied here, such as the vortex-to-biased-density and biased-density-to-Meissner-like transitions. Real-space wires are also the implementation most directly described by our local interaction model [c.f. Eq.~\ref{eq:gpe_1}]. Tunnel-coupled one-dimensional Bose gases have already been realised in optical and atom-chip geometries, with coherent inter-wire coupling and the relative phase along the gases measured through matter-wave interference~\cite{hofferberth2007non,chen2011many}. Periodic discrete boundaries are more demanding, but cylindrical and toroidal arrays of coupled wires have been proposed using Laguerre-Gaussian optical potentials \cite{budich2017coupled}.

An alternative experimental route would be to encode the wire index $\lambda$ in a synthetic dimension, where a set of internal degrees of freedom are externally-coupled together and re-interpreted as sites along an artifical (discrete) direction~\cite{ozawa2017synthetic}. In fact, synthetic dimensions based on internal atomic states~\cite{stuhl2015visualizing, mancini2015observation, celi2014synthetic, chalopin2020probing} have already explored the single-particle physics of coupled wire models~\cite{chalopin2020probing,bouhiron2024realization,li2022bec}. Periodic synthetic boundary conditions can also straightforwardly realised along the synthetic dimensions in these schemes by adding an extra coupling between different internal states~\cite{li2022bec, bouhiron2024realization}. Previously, mean-field theory has  predicted interacting vortex-like lattices for the scheme of Ref.~\cite{chalopin2020probing}, taking into account experimental details such as the specific number of internal states, non-uniform hopping amplitudes and additional light shifts. Our work has instead focused on varying system sizes in the ideal model, but it would be straightforward in the future to adapt our results to include such details to describe atomic state schemes with varying numbers of sublevels. 

 It has also been previously theoretically proposed to realize a coupled wire model in a synthetic dimension of atomic trap states~\cite{price2017synthetic, salerno2019quantized, oliver2023bloch, reid2026phases}. In this approach the inter-wire coupling is engineered by shaking the trap, while the Peierls phase can be tuned by varying the modulation phase along a real continuous spatial dimension. This scheme has been used experimentally to realize a  long synthetic dimension, corresponding of tens of sites~\cite{oliver2023bloch}. Interestingly, this scheme also naturally has an unusual long-range interactions along the synthetic dimension, due to the non-zero spatial overlap between different atomic trap states. As previously shown for the Harper-Hofstadter model, this can lead to the emergence of new types of phases, such as the so-called Meissner stripe phases~\cite{reid2026phases}. In the future, it would therefore be interesting to explore how the above coupled-wire physics extends to this regime. 

Going further, the strong dependence of the vortex structures on system size and boundary conditions also raises questions about their dynamics and stability. A natural next step would be to study the collective modes of the different phases as well as the motion and interaction of vortices following a quench. It would also be interesting to extend this work to higher dimensions, such as a 4D discrete-continuous quantum Hall model~\cite{bouhiron2024realization}, as interesting vortex structures have been predicted in 4D superfluids~\cite{mccanna2021superfluid,mccanna2024curved, mccanna2024curved2, middleton2024interactions}. Finally, an important longer-term question is how the vortex structures found here evolve beyond the weakly interacting mean-field regime. Increasing the interaction strength may lead to a melting of the vortex lattice  and the emergence of correlated states related to quantum Hall physics~\cite{budich2017coupled}.

\section*{Acknowledgements} 
We thank Martin Long for helpful discussions. This work is supported by the Royal Society via grants
URF\textbackslash R\textbackslash221004 and RGF\textbackslash{}EA\textbackslash{}180121, by the Engineering and Physical Sciences Research Council [grant numbers EP/W016141/1, EP/Y01510X/1 and UKRI2226], the CDT
in Topological Design [Grant No. EP/S02297X/1] and the Leverhume Trust [Grant RPG-2024-124]. The computations described in this paper were performed using the University of Birmingham’s Bluebear HPC service, which provides a High Performance Computing service to the University’s research community. \\

%\section*{Data Availability} 

\bibliography{apssamp}% Produces the bibliography via BibTeX.

@PREAMBLE{
 "\providecommand{\noopsort}[1]{}" 
 # "\providecommand{\singleletter}[1]{#1}%" 
}

@article{chen2024strongly,
  title={Strongly interacting Rydberg atoms in synthetic dimensions with a magnetic flux},
  author={Chen, Tao and Huang, Chenxi and Velkovsky, Ivan and Hazzard, Kaden RA and Covey, Jacob P and Gadway, Bryce},
  journal={Nature communications},
  volume={15},
  number={1},
  pages={2675},
  year={2024},
  publisher={Nature Publishing Group UK London}
}

@article{yu2025comprehensive,
  title={Comprehensive review on developments of synthetic dimensions},
  author={Yu, Danying and Song, Wange and Wang, Luojia and Srikanth, Rohith and Kaushik Sridhar, Sashank and Chen, Tao and Huang, Chenxi and Li, Guangzhen and Qiao, Xin and Wu, Xiaoxiong and others},
  journal={Photonics Insights},
  volume={4},
  number={2},
  pages={R06--R06},
  year={2025},
  publisher={Society of Photo-Optical Instrumentation Engineers}
}

@article{kanungo2022realizing,
  title={Realizing topological edge states with Rydberg-atom synthetic dimensions},
  author={Kanungo, Soumya K and Whalen, Joseph D and Lu, Yi and Yuan, M and Dasgupta, S and Dunning, FB and Hazzard, KRA and Killian, Tom C},
  journal={Nature communications},
  volume={13},
  number={1},
  pages={972},
  year={2022},
  publisher={Nature Publishing Group UK London}
}

@article{raghuram2026probing,
  title={Probing topological edge states in a molecular synthetic dimension},
  author={Raghuram, Adarsh P and Blondell, Francesca M and Mortlock, Jonathan M and Maddox, Benjamin P and Dasgupta, Sohail and Middleton-Spencer, Holly AJ and Hazzard, Kaden RA and Price, Hannah M and Gregory, Philip D and Cornish, Simon L},
  journal={arXiv preprint arXiv:2604.00745},
  year={2026}
}

@article{reid2026phases,
  title={Phases of interacting bosons in a hybrid Harper-Hofstadter system with a synthetic dimension of harmonic trap states},
  author={Reid, David G and Middleton-Spencer, Holly AJ and Salerno, Grazia and Goldman, Nathan and Price, Hannah M},
  journal={arXiv preprint arXiv:2602.21108},
  year={2026}
}

@article{leonard2023realization,
  title={Realization of a fractional quantum Hall state with ultracold atoms},
  author={L{\'e}onard, Julian and Kim, Sooshin and Kwan, Joyce and Segura, Perrin and Grusdt, Fabian and Repellin, C{\'e}cile and Goldman, Nathan and Greiner, Markus},
  journal={Nature},
  volume={619},
  number={7970},
  pages={495--499},
  year={2023},
  publisher={Nature Publishing Group UK London}
}

@article{serafini2017vortex,
  title={Vortex reconnections and rebounds in trapped atomic Bose-Einstein condensates},
  author={Serafini, Simone and Galantucci, Luca and Iseni, Elena and Bienaim{\'e}, Tom and Bisset, Russell N and Barenghi, Carlo F and Dalfovo, Franco and Lamporesi, Giacomo and Ferrari, Gabriele},
  journal={Physical Review X},
  volume={7},
  number={2},
  pages={021031},
  year={2017},
  publisher={APS}
}

@article{seman2010three,
  title={Three-vortex configurations in trapped Bose-Einstein condensates},
  author={Seman, JA and Henn, EAL and Haque, M and Shiozaki, RF and Ramos, ERF and Caracanhas, M and Castilho, P and Castelo Branco, C and Tavares, PES and Poveda-Cuevas, FJ and others},
  journal={Physical Review A—Atomic, Molecular, and Optical Physics},
  volume={82},
  number={3},
  pages={033616},
  year={2010},
  publisher={APS}
}

@article{rosenbusch2002dynamics,
  title={Dynamics of a single vortex line in a Bose-Einstein condensate},
  author={Rosenbusch, Peter and Bretin, Vincent and Dalibard, Jean},
  journal={Physical review letters},
  volume={89},
  number={20},
  pages={200403},
  year={2002},
  publisher={APS}
}

@article{kasamatsu2005three,
  title={Three-dimensional dynamics of vortex-lattice formation in Bose-Einstein condensates},
  author={Kasamatsu, Kenichi and Machida, Masahiko and Sasa, Narimasa and Tsubota, Makoto},
  journal={Physical Review A—Atomic, Molecular, and Optical Physics},
  volume={71},
  number={6},
  pages={063616},
  year={2005},
  publisher={APS}
}

@article{aftalion2003three,
  title={Three-dimensional vortex configurations in a rotating Bose-Einstein condensate},
  author={Aftalion, Amandine and Danaila, Ionut},
  journal={Physical Review A},
  volume={68},
  number={2},
  pages={023603},
  year={2003},
  publisher={APS}
}

@article{tsubota2002vortex,
  title={Vortex lattice formation in a rotating Bose-Einstein condensate},
  author={Tsubota, Makoto and Kasamatsu, Kenichi and Ueda, Masahito},
  journal={Physical Review A},
  volume={65},
  number={2},
  pages={023603},
  year={2002},
  publisher={APS}
}

@article{mccanna2024curved,
  title={Curved vortex surfaces in four-dimensional superfluids. I. Unequal-frequency double rotations},
  author={McCanna, Ben and Price, Hannah M},
  journal={Physical Review A},
  volume={110},
  number={1},
  pages={013325},
  year={2024},
  publisher={APS}
}

@article{mccanna2024curved2,
  title={Curved vortex surfaces in four-dimensional superfluids. II. Equal-frequency double rotations},
  author={McCanna, Ben and Price, Hannah M},
  journal={Physical Review A},
  volume={110},
  number={1},
  pages={013326},
  year={2024},
  publisher={APS}
}

@article{mccanna2021superfluid,
  title={Superfluid vortices in four spatial dimensions},
  author={McCanna, Ben and Price, Hannah M},
  journal={Physical Review Research},
  volume={3},
  number={2},
  pages={023105},
  year={2021},
  publisher={APS}
}

@article{middleton2024interactions,
  title={Interactions and Reconnections of Four-Dimensional Quantum Vortices},
  author={Middleton-Spencer, HAJ and McCanna, B and Proment, D and Price, HM},
  journal={arXiv preprint arXiv:2411.07943},
  year={2024}
}

@article{PhysRevB.99.035130,
  title = {Quantum Hall hierarchy from coupled wires},
  author = {Fuji, Yohei and Furusaki, Akira},
  journal = {Phys. Rev. B},
  volume = {99},
  issue = {3},
  pages = {035130},
  numpages = {34},
  year = {2019},
  month = {Jan},
  publisher = {American Physical Society},
  doi = {10.1103/PhysRevB.99.035130},
  url = {https://link.aps.org/doi/10.1103/PhysRevB.99.035130}
}

@article{PhysRevB.91.245139,
  title = {Chiral spin liquids in arrays of spin chains},
  author = {Gorohovsky, Gregory and Pereira, Rodrigo G. and Sela, Eran},
  journal = {Phys. Rev. B},
  volume = {91},
  issue = {24},
  pages = {245139},
  numpages = {12},
  year = {2015},
  month = {Jun},
  publisher = {American Physical Society},
  doi = {10.1103/PhysRevB.91.245139},
  url = {https://link.aps.org/doi/10.1103/PhysRevB.91.245139}
}

@article{PhysRevB.89.085101,
  title = {From Luttinger liquid to non-Abelian quantum Hall states},
  author = {Teo, Jeffrey C. Y. and Kane, C. L.},
  journal = {Phys. Rev. B},
  volume = {89},
  issue = {8},
  pages = {085101},
  numpages = {22},
  year = {2014},
  month = {Feb},
  publisher = {American Physical Society},
  doi = {10.1103/PhysRevB.89.085101},
  url = {https://link.aps.org/doi/10.1103/PhysRevB.89.085101}
}

@article{meng2015coupled,
  title={Coupled-wire construction of chiral spin liquids},
  author={Meng, Tobias and Neupert, Titus and Greiter, Martin and Thomale, Ronny},
  journal={Physical Review B},
  volume={91},
  number={24},
  pages={241106},
  year={2015},
  publisher={APS}
}

@article{williams2010observation,
  title={Observation of Vortex Nucleation in a Rotating Two-Dimensional Lattice of Bose-Einstein Condensates},
  author={Williams, RA and Al-Assam, S and Foot, CJ},
  journal={Physical review letters},
  volume={104},
  number={5},
  pages={050404},
  year={2010},
  publisher={APS}
}

@article{cooper2008rapidly,
  title={Rapidly rotating atomic gases},
  author={Cooper, Nigel R},
  journal={Advances in Physics},
  volume={57},
  number={6},
  pages={539--616},
  year={2008},
  publisher={Taylor \& Francis}
}

@article{cooper2005vortex,
  title={Vortex lattices in rotating atomic Bose gases with dipolar interactions},
  author={Cooper, NR and Rezayi, EH and Simon, SH},
  journal={Physical review letters},
  volume={95},
  number={20},
  pages={200402},
  year={2005},
  publisher={APS}
}

@article{coddington2003observation,
  title={Observation of Tkachenko oscillations in rapidly rotating Bose-Einstein condensates},
  author={Coddington, Ian and Engels, Peter and Schweikhard, Volker and Cornell, Eric A},
  journal={Physical review letters},
  volume={91},
  number={10},
  pages={100402},
  year={2003},
  publisher={APS}
}

@article{tung2006observation,
  title={Observation of vortex pinning in Bose-Einstein condensates},
  author={Tung, S and Schweikhard, V and Cornell, Eric A},
  journal={Physical review letters},
  volume={97},
  number={24},
  pages={240402},
  year={2006},
  publisher={APS}
}

@article{abrikosov2004nobel,
  title={Nobel Lecture: Type-II superconductors and the vortex lattice},
  author={Abrikosov, Aleksej A},
  journal={Reviews of modern physics},
  volume={76},
  number={3},
  pages={975--979},
  year={2004},
  publisher={APS}
}

@article{PhysRevLett.88.036401,
  title = {Fractional Quantum Hall Effect in an Array of Quantum Wires},
  author = {Kane, C. L. and Mukhopadhyay, Ranjan and Lubensky, T. C.},
  journal = {Phys. Rev. Lett.},
  volume = {88},
  issue = {3},
  pages = {036401},
  numpages = {4},
  year = {2002},
  month = {Jan},
  publisher = {American Physical Society},
  doi = {10.1103/PhysRevLett.88.036401},
  url = {https://link.aps.org/doi/10.1103/PhysRevLett.88.036401}
}

@article{powell2011bogoliubov,
  title={Bogoliubov theory of interacting bosons on a lattice in a synthetic magnetic field},
  author={Powell, Stephen and Barnett, Ryan and Sensarma, Rajdeep and Das Sarma, Sankar},
  journal={Physical Review A—Atomic, Molecular, and Optical Physics},
  volume={83},
  number={1},
  pages={013612},
  year={2011},
  publisher={APS}
}

@article{schafer2020tools,
  title={Tools for quantum simulation with ultracold atoms in optical lattices},
  author={Sch{\"a}fer, Florian and Fukuhara, Takeshi and Sugawa, Seiji and Takasu, Yosuke and Takahashi, Yoshiro},
  journal={Nature Reviews Physics},
  volume={2},
  number={8},
  pages={411--425},
  year={2020},
  publisher={Nature Publishing Group UK London}
}

@article{bouhiron2024realization,
  title={Realization of an atomic quantum Hall system in four dimensions},
  author={Bouhiron, Jean-Baptiste and Fabre, Aur{\'e}lien and Liu, Qi and Redon, Quentin and Mittal, Nehal and Satoor, Tanish and Lopes, Raphael and Nascimbene, Sylvain},
  journal={Science},
  volume={384},
  number={6692},
  pages={223--227},
  year={2024},
  publisher={American Association for the Advancement of Science}
}

@article{oliver2023bloch,
  title={Bloch oscillations along a synthetic dimension of atomic trap states},
  author={Oliver, Christopher and Smith, Aaron and Easton, Thomas and Salerno, Grazia and Guarrera, Vera and Goldman, Nathan and Barontini, Giovanni and Price, Hannah M},
  journal={Physical Review Research},
  volume={5},
  number={3},
  pages={033001},
  year={2023},
  publisher={APS}
}

@article{goldman2016topological,
  title={Topological quantum matter with ultracold gases in optical lattices},
  author={Goldman, Nathan and Budich, Jan C and Zoller, Peter},
  journal={Nature Physics},
  volume={12},
  number={7},
  pages={639--645},
  year={2016},
  publisher={Nature Publishing Group UK London}
}

@article{atala2014observation,
  title={Observation of chiral currents with ultracold atoms in bosonic ladders},
  author={Atala, Marcos and Aidelsburger, Monika and Lohse, Michael and Barreiro, Julio T and Paredes, Bel{\'e}n and Bloch, Immanuel},
  journal={Nature Physics},
  volume={10},
  number={8},
  pages={588--593},
  year={2014},
  publisher={Nature Publishing Group UK London}
}

@article{abo2001observation,
  title={Observation of vortex lattices in Bose-Einstein condensates},
  author={Abo-Shaeer, Jamil R and Raman, Chandra and Vogels, Johnny M and Ketterle, Wolfgang},
  journal={Science},
  volume={292},
  number={5516},
  pages={476--479},
  year={2001},
  publisher={American Association for the Advancement of Science}
}

@article{orignac2001meissner,
  title={Meissner effect in a bosonic ladder},
  author={Orignac, E and Giamarchi, T},
  journal={Physical Review B},
  volume={64},
  number={14},
  pages={144515},
  year={2001},
  publisher={APS}
}

@article{mancini2015observation,
  title={Observation of chiral edge states with neutral fermions in synthetic Hall ribbons},
  author={Mancini, Marco and Pagano, Guido and Cappellini, Giacomo and Livi, Lorenzo and Rider, Marie and Catani, Jacopo and Sias, Carlo and Zoller, Peter and Inguscio, Massimo and Dalmonte, Marcello and others},
  journal={Science},
  volume={349},
  number={6255},
  pages={1510--1513},
  year={2015},
  publisher={American Association for the Advancement of Science}
}

@article{stuhl2015visualizing,
  title={Visualizing edge states with an atomic Bose gas in the quantum Hall regime},
  author={Stuhl, BK and Lu, H-I and Aycock, LM and Genkina, D and Spielman, IB},
  journal={Science},
  volume={349},
  number={6255},
  pages={1514--1518},
  year={2015},
  publisher={American Association for the Advancement of Science}
}

@article{price2017synthetic,
  title={Synthetic dimensions for cold atoms from shaking a harmonic trap},
  author={Price, Hannah M and Ozawa, Tomoki and Goldman, Nathan},
  journal={Physical Review A},
  volume={95},
  number={2},
  pages={023607},
  year={2017},
  publisher={APS}
}

@article{PhysRevA.111.033301,
  title = {Topological chiral edge states in a synthetic dimension of atomic trap states},
  author = {Reid, David G. and Oliver, Christopher and Regan, Patrick and Smith, Aaron and Easton, Thomas and Salerno, Grazia and Barontini, Giovanni and Goldman, Nathan and Price, Hannah M.},
  journal = {Phys. Rev. A},
  volume = {111},
  issue = {3},
  pages = {033301},
  numpages = {20},
  year = {2025},
  month = {Mar},
  publisher = {American Physical Society},
  doi = {10.1103/PhysRevA.111.033301},
}

@article{PhysRevB.101.045102,
  title = {Variational Bethe ansatz approach for dipolar one-dimensional bosons},
  author = {De Palo, S. and Citro, R. and Orignac, E.},
  journal = {Phys. Rev. B},
  volume = {101},
  issue = {4},
  pages = {045102},
  numpages = {12},
  year = {2020},
  month = {Jan},
  publisher = {American Physical Society},
  doi = {10.1103/PhysRevB.101.045102},
}

@article{Clark_2006,
   title={Off-Diagonal Long-Range Order in Solid $^{4}${He}},
   volume={96},
   ISSN={1079-7114},
   url={http://dx.doi.org/10.1103/PhysRevLett.96.105302},
   DOI={10.1103/physrevlett.96.105302},
   number={10},
   journal={Physical Review Letters},
   publisher={American Physical Society (APS)},
   author={Clark, Bryan K. and Ceperley, D. M.},
   year={2006},
   month=mar }

@article{
doi:10.1126/science.235.4793.1196,
author = {P. W. Anderson },
title = {The Resonating Valence Bond State in {La$_2$CuO$_4$} and Superconductivity},
journal = {Science},
volume = {235},
number = {4793},
pages = {1196-1198},
year = {1987},
doi = {10.1126/science.235.4793.1196}
}

@article{10.1093/comjnl/13.3.317,
    author = {Fletcher, R.},
    title = {A new approach to variable metric algorithms},
    journal = {The Computer Journal},
    volume = {13},
    number = {3},
    pages = {317-322},
    year = {1970},
    month = {01},
    issn = {0010-4620},
    doi = {10.1093/comjnl/13.3.317},
}

@article{PhysRevLett.50.1395,
  title = {Anomalous Quantum Hall Effect: An Incompressible Quantum Fluid with Fractionally Charged Excitations},
  author = {Laughlin, R. B.},
  journal = {Phys. Rev. Lett.},
  volume = {50},
  issue = {18},
  pages = {1395--1398},
  numpages = {0},
  year = {1983},
  month = {May},
  publisher = {American Physical Society},
  doi = {10.1103/PhysRevLett.50.1395},
}

@article{
doi:10.1126/sciadv.adj0360,
author = {Christopher Oliver  and Sebabrata Mukherjee  and Mikael C. Rechstman  and Iacopo Carusotto  and Hannah M. Price },
title = {Artificial gauge fields in the $t-z$ mapping for optical pulses: Spatiotemporal wave packet control and quantum Hall physics},
journal = {Science Advances},
volume = {9},
number = {42},
year = {2023},
eprint = {https://www.science.org/doi/pdf/10.1126/sciadv.adj0360}}

@article{budich2017coupled,
  title={Coupled atomic wires in a synthetic magnetic field},
  author={FIX!, Budich, JC and Elben, A and {L}{{a}}cki, M and Sterdyniak, Antoine and Baranov, MA and Zoller, P},
  journal={Physical Review A},
  volume={95},
  number={4},
  pages={043632},
  year={2017},
  publisher={APS}
}

@article{celi2014synthetic,
  title={Synthetic gauge fields in synthetic dimensions},
  author={Celi, Alessio and Massignan, Pietro and Ruseckas, Julius and Goldman, Nathan and Spielman, Ian B and Juzeli{\=u}nas, G and Lewenstein, M},
  journal={Physical review letters},
  volume={112},
  number={4},
  pages={043001},
  year={2014},
  publisher={APS}
}

@article{chalopin2020probing,
  title={Probing chiral edge dynamics and bulk topology of a synthetic Hall system},
  author={Chalopin, Thomas and Satoor, Tanish and Evrard, Alexandre and Makhalov, Vasiliy and Dalibard, Jean and Lopes, Raphael and Nascimbene, Sylvain},
  journal={Nature Physics},
  volume={16},
  number={10},
  pages={1017--1021},
  year={2020},
  publisher={Nature Publishing Group UK London}
}

@article{meng2020coupled,
  title={Coupled-wire constructions: a Luttinger liquid approach to topology},
  author={Meng, Tobias},
  journal={The European Physical Journal Special Topics},
  volume={229},
  number={4},
  pages={527--543},
  year={2020},
  publisher={Springer}
}

@article{ozawa2019topological,
  title={Topological quantum matter in synthetic dimensions},
  author={Ozawa, Tomoki and Price, Hannah M},
  journal={Nature Reviews Physics},
  volume={1},
  number={5},
  pages={349--357},
  year={2019},
  publisher={Nature Publishing Group UK London}
}

@article{Oreg2014Helical,
  title = {Fractional helical liquids in quantum wires},
  author = {Oreg, Yuval and Sela, Eran and Stern, Ady},
  journal = {Phys. Rev. B},
  volume = {89},
  issue = {11},
  pages = {115402},
  numpages = {8},
  year = {2014},
  month = {Mar},
  publisher = {American Physical Society},
  doi = {10.1103/PhysRevB.89.115402},
}

@article{Oreg2019Fractional,
  title = {Fractional Conductance in Strongly Interacting 1D Systems},
  author = {Shavit, Gal and Oreg, Yuval},
  journal = {Phys. Rev. Lett.},
  volume = {123},
  issue = {3},
  pages = {036803},
  numpages = {6},
  year = {2019},
  month = {Jul},
  publisher = {American Physical Society},
  doi = {10.1103/PhysRevLett.123.036803},
  url = {https://link.aps.org/doi/10.1103/PhysRevLett.123.036803}
}

@article{Wei2014TwoLegBosons,
  title = {Theory of bosons in two-leg ladders with large magnetic fields},
  author = {Wei, Ran and Mueller, Erich J.},
  journal = {Phys. Rev. A},
  volume = {89},
  issue = {6},
  pages = {063617},
  numpages = {6},
  year = {2014},
  month = {Jun},
  publisher = {American Physical Society},
  doi = {10.1103/PhysRevA.89.063617},
  url = {https://link.aps.org/doi/10.1103/PhysRevA.89.063617}
}

@article{byers1961theoretical,
  title={Theoretical considerations concerning quantized magnetic flux in superconducting cylinders},
  author={Byers, N and Yang, CN},
  journal={Physical review letters},
  volume={7},
  number={2},
  pages={46},
  year={1961},
  publisher={APS}
}

@article{li2012tricriticality,
  title={Quantum tricriticality and phase transitions in spin-orbit coupled Bose-Einstein condensates},
  author={Li, Yun and Pitaevskii, Lev P and Stringari, Sandro},
  journal={Physical review letters},
  volume={108},
  number={22},
  pages={225301},
  year={2012},
  publisher={APS}
}

@article{li2013superstripes,
  title = {Superstripes and the Excitation Spectrum of a Spin-Orbit-Coupled Bose-Einstein Condensate},
  author = {Li, Yun and Martone, Giovanni I. and Pitaevskii, Lev P. and Stringari, Sandro},
  journal = {Phys. Rev. Lett.},
  volume = {110},
  issue = {23},
  pages = {235302},
  numpages = {5},
  year = {2013},
  month = {Jun},
  publisher = {American Physical Society},
  doi = {10.1103/PhysRevLett.110.235302},
  url = {https://link.aps.org/doi/10.1103/PhysRevLett.110.235302}
}

@article{li2017stripe,
  title={A stripe phase with supersolid properties in spin--orbit-coupled Bose--Einstein condensates},
  author={Li, Jun-Ru and Lee, Jeongwon and Huang, Wujie and Burchesky, Sean and Shteynas, Boris and Top, Furkan {\c{C}}a{\u{g}}r{\i} and Jamison, Alan O and Ketterle, Wolfgang},
  journal={Nature},
  volume={543},
  number={7643},
  pages={91--94},
  year={2017},
  publisher={Nature Publishing Group UK London}
}

@article{ji2014experimental,
  title={Experimental determination of the finite-temperature phase diagram of a spin--orbit coupled Bose gas},
  author={Ji, Si-Cong and Zhang, Jin-Yi and Zhang, Long and Du, Zhi-Dong and Zheng, Wei and Deng, You-Jin and Zhai, Hui and Chen, Shuai and Pan, Jian-Wei},
  journal={Nature physics},
  volume={10},
  number={4},
  pages={314--320},
  year={2014},
  publisher={Nature Publishing Group UK London}
}

@article{qiu2021stripe,
  title={Stripe and junction-vortex phases in linearly coupled Bose-Einstein condensates},
  author={Qiu, Haibo and Zhang, Dengling and Mateo, Antonio Mu{\~n}oz},
  journal={Physical Review A},
  volume={103},
  number={2},
  pages={023316},
  year={2021},
  publisher={APS}
}

@article{citro2020spectral,
  title={Spectral function of a boson ladder in an artificial gauge field},
  author={Citro, Roberta and De Palo, Stefania and Victorin, Nicolas and Minguzzi, Anna and Orignac, Edmond},
  journal={Condensed Matter},
  volume={5},
  number={1},
  pages={15},
  year={2020},
  publisher={MDPI}
}

@article{cabedo2020effective,
  title={Effective triangular ladders with staggered flux from spin-orbit coupling in 1D optical lattices},
  author={Cabedo, Josep and Claramunt, Joan and Mompart, Jordi and Ahufinger, Veronica and Celi, Alessio},
  journal={The European Physical Journal D},
  volume={74},
  number={6},
  pages={123},
  year={2020},
  publisher={Springer}
}

@article{hofferberth2007non,
  title={Non-equilibrium coherence dynamics in one-dimensional Bose gases},
  author={Hofferberth, S and Lesanovsky, Igor and Fischer, B and Schumm, Thorsten and Schmiedmayer, J{\"o}rg},
  journal={Nature},
  volume={449},
  number={7160},
  pages={324--327},
  year={2007},
  publisher={Nature Publishing Group UK London}
}

@article{chen2011many,
  title={Many-body Landau--Zener dynamics in coupled one-dimensional Bose liquids},
  author={Chen, Yu-Ao and Huber, Sebastian D and Trotzky, Stefan and Bloch, Immanuel and Altman, Ehud},
  journal={Nature Physics},
  volume={7},
  number={1},
  pages={61--67},
  year={2011},
  publisher={Nature Publishing Group UK London}
}

@article{li2022bec,
  title={Bose-Einstein condensate on a synthetic topological Hall cylinder},
  author={Li, Chuan-Hsun and Yan, Yangqian and Feng, Shih-Wen and Choudhury, Sayan and Blasing, David B and Zhou, Qi and Chen, Yong P},
  journal={PRX Quantum},
  volume={3},
  number={1},
  pages={010316},
  year={2022},
  publisher={APS}
}

@article{salerno2019quantized,
  title={Quantized Hall conductance of a single atomic wire: a proposal based on synthetic dimensions},
  author={Salerno, Grazia and Price, Hannah M and Lebrat, Martin and H{\"a}usler, Samuel and Esslinger, Tilman and Corman, Laura and Brantut, J-P and Goldman, Nathan},
  journal={Physical Review X},
  volume={9},
  number={4},
  pages={041001},
  year={2019},
  publisher={APS}
}

@article{harper1955UniformBfield,
doi = {10.1088/0370-1298/68/10/304},
url = {https://doi.org/10.1088/0370-1298/68/10/304},
year = {1955},
month = {oct},
publisher = {},
volume = {68},
number = {10},
pages = {874},
author = {P G Harper},
title = {Single Band Motion of Conduction Electrons in a Uniform Magnetic Field},
journal = {Proceedings of the Physical Society. Section A}
}

@article{piraud2015vortex,
  title={Vortex and Meissner phases of strongly interacting bosons on a two-leg ladder},
  author={Piraud, Marie and Heidrich-Meisner, Fabian and McCulloch, Ian P and Greschner, Sebastian and Vekua, Temo and Schollwoeck, Ulrich},
  journal={Physical Review B},
  volume={91},
  number={14},
  pages={140406},
  year={2015},
  publisher={APS}
}

@article{kelecs2015mott,
  title={Mott transition in a two-leg Bose-Hubbard ladder under an artificial magnetic field},
  author={Kele{\c{s}}, Ahmet and Oktel, M{\"O}},
  journal={Physical Review A},
  volume={91},
  number={1},
  pages={013629},
  year={2015},
  publisher={APS}
}

@article{qiao2021quantum,
  title={Quantum phases of interacting bosons on biased two-leg ladders with magnetic flux},
  author={Qiao, Xin and Zhang, Xiao-Bo and Jian, Yue and Zhang, Ai-Xia and Yu, Zi-Fa and Xue, Ju-Kui},
  journal={Physical Review A},
  volume={104},
  number={5},
  pages={053323},
  year={2021},
  publisher={APS}
}

@article{uchino2015population,
  title={Population-imbalance instability in a Bose-Hubbard ladder in the presence of a magnetic flux},
  author={Uchino, Shun and Tokuno, Akiyuki},
  journal={Physical Review A},
  volume={92},
  number={1},
  pages={013625},
  year={2015},
  publisher={APS}
}

@article{natu2015bosons,
  title={Bosons with long-range interactions on two-leg ladders in artificial magnetic fields},
  author={Natu, Stefan S},
  journal={Physical Review A},
  volume={92},
  number={5},
  pages={053623},
  year={2015},
  publisher={APS}
}

@article{kolovsky2017bogoliubov,
  title={Bogoliubov depletion of the fragmented condensate in the bosonic flux ladder},
  author={Kolovsky, Andrey R},
  journal={Physical Review A},
  volume={95},
  number={3},
  pages={033622},
  year={2017},
  publisher={APS}
}

@article{oliver2023artificial,
  title={Artificial gauge fields in the t-z mapping for optical pulses: Spatiotemporal wave packet control and quantum Hall physics},
  author={Oliver, Christopher and Mukherjee, Sebabrata and Rechstman, Mikael C and Carusotto, Iacopo and Price, Hannah M},
  journal={Science Advances},
  volume={9},
  number={42},
  pages={eadj0360},
  year={2023},
  publisher={American Association for the Advancement of Science}
}

@article{ozawa2021artificial,
  title={Artificial magnetic field for synthetic quantum matter without dynamical modulation},
  author={Ozawa, Tomoki},
  journal={Physical Review A},
  volume={103},
  number={3},
  pages={033318},
  year={2021},
  publisher={APS}
}

@article{ozawa2017synthetic,
  title={Synthetic dimensions with magnetic fields and local interactions in photonic lattices},
  author={Ozawa, Tomoki and Carusotto, Iacopo},
  journal={Physical review letters},
  volume={118},
  number={1},
  pages={013601},
  year={2017},
  publisher={APS}
}

@article{price2020synthetic,
  title={Synthetic dimensions and topological chiral currents in mesoscopic rings},
  author={Price, Hannah M and Ozawa, Tomoki and Schomerus, Henning},
  journal={Physical Review Research},
  volume={2},
  number={3},
  pages={032017},
  year={2020},
  publisher={APS}
}

\end{document}